\documentstyle[preprint,eqsecnum,aps,floats]{revtex}
\tighten
\begin{document}
\preprint{SU-GP-97/10-1}
\input{epsf.tex}
\draft
\def\be{\begin{equation}} 
\def\ba{\begin{eqnarray}} 
\def\ee{\end{equation}} 
\def\ea{\end{eqnarray}} 
\def\g{\gamma}
\def\Cyl{{\rm Cyl}}
\def\T{{\cal T}}
\def\C{{\cal T}}
\def\H{{\cal H}}
\def\inv{{\rm inv}}
\def\a{\alpha}
\def\e{\epsilon}
\def\g{\gamma}
\def\h{\hat{h}}
\def\hH{\hat{H}}
\def\Diff{{\rm Diff}}
\def\phys{{\rm phys}}
\def\big{{\rm big}}
\def\k{\kappa}
\def\D{\cal D}
\def\s{{\cal s}}
\def\av{{\rm diff}}
\title{Loop constraints: A habitat and their algebra}

\author{Jerzy Lewandowski\footnote{A Humboldt fellow, on leave from 
Instytut Fizyki Teoretycznej,
Uniwersytet Warszawski, ul. Ho\.za 69, 00-681 Warszawa, Poland} \\{\em
Max-Planck-Institut f\"ur Gravitationsphysik,
Schlaatzweg 1,
D-14473 Potsdam, Germany }}
\author{and}
\author{Donald Marolf
\\{\em Physics Department, Syracuse University, Syracuse, NY 13244-1130}}
\date{\today}
\maketitle

\begin{abstract}
This work introduces a new space $\T'_*$ of `vertex-smooth' states for use
in the loop approach to quantum gravity.  Such states provide
a natural domain for Euclidean Hamiltonian constraint operators of the type 
introduced by Thiemann (and using certain ideas of Rovelli and Smolin).
In particular, such operators map
$\T'_*$ into itself, and so are actual operators {\it in} this
space.  Their commutator can be computed on $\T'_*$
and compared with the classical hypersurface deformation algebra.
Although the classical Poisson bracket of Hamiltonian constraints
yields an inverse
metric times an infinitesimal diffeomorphism generator, and despite the
fact that the diffeomorphism generator has a well-defined non-trivial 
action on $\T'_*$, the commutator of quantum constraints
vanishes identically for a large class of proposals.
 
\end{abstract}

\section{Introduction.}    

Within the loop-based approach to quantum gravity, there are now a number
of proposals for the Hamiltonian constraint \cite{I,II,Lee,DPR}. 
Most of these are modifications of Thiemann's proposal \cite{I}, 
and in particular make use of an observation by Rovelli and Smolin \cite{CL}
that certain limits of operators can be taken on diffeomorphism invariant
states.  One would like to
test any proposal for the quantum constraints of gravity
in a variety of ways.  Below, we consider the proposals for 
Euclidean quantum gravity, computing the constraint algebras for each
and comparing them to the classical
hypersurface deformation algebra of \cite{BD,TR}.

Although the constraint algebra has been studied at a more heuristic
level and in a less well-defined context \cite{BB,GGP},
an actual computation has until now been impossible for the
proposals of \cite{I,II,DPR} due to the subtle
way in which these works construct their constraints.  The proposals
follow the ideas of \cite{CL} and 
define the constraints only when acting on a space ${\T'_{Diff}}$ of
`diffeomorphism invariant'
states.  This is because they employ a limiting
procedure which does not converge on a general state.
However, because a typical Hamiltonian constraint has 
nonvanishing commutator with the diffeomorphism constraint, the
action of such a constraint takes a diffeomorphism invariant state
to a state that is {\it not} diffeomorphism invariant.  The ranges of the
proposed constraint operators are therefore not contained in their domains and
it is not possible to apply two of them 
in succession, or to directly compute a commutator.  It is important
to note that what we have in mind differs from the ``anomaly-free''
calculation of \cite{I} in that we wish to commute the so-called
`unregulated' or `regulator independent' operators, whereas  \cite{I}
studied the commutator of regulator dependent constraints.

The main result of this paper is
that the limiting procedures of \cite{I,CL} in fact converge on a larger
space $\T'_* \supset \T'_{Diff}$,
which we shall call the space of `vertex-smooth states.'  Thus, 
the proposed operators extend naturally to $\T'_*$.  Furthermore, 
$\T'_*$ is mapped into itself by all of the proposed constraints.
As a result, the proposals define constraint
operators {\it within} the space $\T'_*$, 
and products and commutators
of such operators are well defined in this
space.  

Let us recall that, classically, the Poisson bracket
of two Euclidean 
Hamiltonian constraints is an inverse metric $q^{ab}$ times an infinitesimal
diffeomorphism generator $C_b$:
\be 
\label{alg1}
\{H(N),H(M)\} = \int (MN_a - NM_a) C_b q^{ab}. 
\ee
We will see that a
generic element of $\T'_*$ is not
annihilated by the diffeomorphism generator\footnote{The infinitesimal
diffeomorphism generators are in fact well defined on $\T'_*$, a fact
first pointed out to the authors by Jos\'e Mour\~ao.} Indeed, the action of
the diffeomorphism group on $\T'_*$ provides a faithful representation.
In addition, $\T'_*$ contains the entire space of solutions
to the constraints discussed in \cite{I,II} -- presumably, the entire
space of physical states in these proposals.  It would therefore be
a great surprise if the inverse metric was degenerate on this space.
Nevertheless, we find that the commutator of two Hamiltonian constraints
vanishes identically on $\T'_*$
for a large class of proposals.  More will be said about
quantum versions of $\int N_a C_b q^{ab}$ 
in the accompanying paper \cite{GLMP}.

There is in fact a general difficulty in constructing a quantum
version of \ref{alg1} using operators that act on (and preserve)
some subspace of a Hilbert space which contains diffeomorphism 
invariant states.
With a few natural assumptions, we shown in the Appendix that, in such a case,
every diffeomorphism invariant state in
domain of the Hamiltonian operators
must be annihilated by the Hamiltonians.  It is interesting to note
that our argument 
breaks down if the constraints are rescaled and made into 
minus-half-densities -- a case never considered
in canonical gravity to our knowledge.        

The plan of this paper is as follows.  Section II first establishes 
the context and conventions for our work
and then describes the new space $\T'_*$.  Section III
then describes a general class of `RST-like' operators on $\T'_*$
which includes
many of the (so-called `non-symmetric')
proposals for the Euclidean Hamiltonian constraints.  It also shows
that the commutator of such operators vanishes in general.  In section
IV, we discuss various `symmetrized' operators that have been proposed.
Here, the commutator again tends to vanish and, when it does not, it also
fails to annihilate
diffeomorphism invariant states.  This accounts for all existing
proposals except that of \cite{Lee}, which will be considered in
\cite{GLMP}.  We end with a brief discussion in section V.

\section{A new space: the vertex-smooth states}

This section introduces the new space $\T'_*$ of `vertex-smooth states' 
which will allow us to compute constraint algebras.  Section IIA
sets the framework for our discussion and establishes notation and
conventions.  Section IIB then describes the vertex-smooth states.  
We save the demonstration that $\T'_*$ provides a natural habitat for
RST-like constraints for a later section, after the constraints themselves
have been introduced.

\subsection{Preliminaries}
We now take a few moments to fix our context and conventions before
introducing the new space.  We recall that standard constructions
\cite{AI,MM,ALmeasure,Bameasure,ALMMT}
of the space of generalized connections   make 
use of an analytic structure on the three manifold $\Sigma$. They 
were generalized   by Baez and Sawin \cite{BS,BSII} to the smooth category,
however the notion of the spin-network  has not been completely
successfully defined in 
that case. On the one hand, the definition of Hamiltonian constraints 
given in \cite{I} 
requires the action of smooth, rather then analytic diffeomorphisms but, on 
the other hand,  
the construction of the  diffeomorphism invariant states of \cite{ALMMT} 
makes use of the spin-networks. Merging these two features
requires some care. The  Hilbert space  we desire is constructed without 
invoking an analytic  structure but it is only a  subspace of that of 
\cite{BS}; in fact, it is the subspace studied in \cite{BSII}. 


In \cite{AI}, a space of `generalized connections' was constructed
using the $C^*$ algebra defined by the traces of
holonomies of a connection along
piecewise smooth closed
curves in $\Sigma$.  The spectrum of this algebra is the
Ashtekar-Isham space $\overline{{\cal A}/{\cal G}}$.  The elements
of this space can be thought of as `distributional' connections for which
the holonomy around any closed
curve is well-defined, but for which such holonomies 
satisfy no continuity properties \cite{ALmeasure}.  This is to be the 
`quantum configuration space,' and quantum states are to be functions 
on this space.
Following \cite{ALdiff} we consider a special set of such functions
associated with graphs embedded in $\Sigma$. By a graph $\gamma$
we mean a finite
set of `edges' (1-dimensional, smooth oriented submanifolds of $\Sigma$
with  a 2-point boundary called `the ends' of an edge) such that 
any two of them intersect, if at all, at only 
one or both ends.  We denote
the set of edges of $\gamma$ by $E(\gamma)$, and the particular subset 
with at least one end at $v$ by $E(\gamma, v)$.  Also 
associated with a graph $\gamma$ is a set of
vertices $V(\gamma)$;
the vertices $V(\gamma)$ are the end points of the edges.

We will say that a function on $\overline{{\cal A}/{\cal G}}$ is `cylindrical
over a graph $\gamma$' if it depends only on the holonomies of the generalized
connection along curves that lie in that particular graph. 
As the graphs we consider are smoothly embedded, every cylindrical 
function over a graph belongs to the Hilbert space described by Baez and 
Sawin \cite{BS}.  As a result, 
there is an inner product on these smooth cylindrical functions, and 
they can be completed to form a Hilbert space ${\cal H}$
which is a proper subspace of the Hilbert space of \cite{BS}. The
construction of \cite{BS} is more general and allows curves to intersect
an infinite number of times, but such cases were not considered
in \cite{I,II,Lee,RR} so we will also exclude them here
(see \cite{LT} for an extension of the theory  to such cases).
The natural action of smooth diffeomorphisms of $\Sigma$ in 
${\cal H}$ is unitary.  For $\varphi \in Diff(\Sigma)$, the action
will be denoted ${\D}_\varphi$, with ${\D}_\varphi |\Gamma \rangle
= |\varphi(\Gamma) \rangle$.

To each graph 
$\g$ one  associates a certain subspace ${\cal H}_\g\subset {\cal H}$ 
in such a manner that ${\cal H}_\gamma$  is orthogonal to ${\cal H}_\gamma'$ 
whenever the ranges of the graphs differ from each other, $R(\g)\not=R(\g')$. 
The Hilbert space ${\cal H}$ has the property that
\be
{\cal H}\ =\ \oplus_{R(\g)}{\cal H}_\g
\ee
where $\oplus$ denotes the direct sum of Hilbert spaces and implies
that that the result should be completed to obtain another Hilbert space.

Given a graph $\g$, the space ${\cal H}_\g$  
can be formed from the associated to $\g$ `spin-network functions' 
\cite{Baez,SN} 
Recall that spin networks $\Gamma$ are smooth cylindrical
functions which are parameterized by triples $(\gamma, j,c)$ where
$\gamma$ ranges
over all graphs embedded in $\Sigma$ and $j,c$ range over certain
lists of `spins' and `contractors' associated with the graph $\gamma$.
The label $j$  assigns a representation of $SU(2)$ to each 
edge  of $\gamma$, while a contractor $c$ assigns an `intertwinor'
to each vertex $v$ in $\gamma$ which ensures that  $\Gamma$ 
is invariant with respect to the gauge transformations.
The intertwinors are linear operators which act in a space
determined by the spins assigned by $j$ to the edges that
intersect at $v$; the reader should consult \cite{Baez,SN} for details.
Now, ${\cal H}_\g$
is the Hilbert completion of the space spanned by all  the spin-network
functions given by all the labels $(\gamma, j,c)$ such that
for every edge $e\in E(\g)$, $j(e)\not=0$.  
It is convenient to use the symbol $V(\Gamma)$ to denote the vertex set
of the underlying graph $\gamma$, and to refer to the vertices of $\gamma$
as vertices of the spin network $\Gamma$. Also, given a graph $\gamma$
or a spin-network $\Gamma$, by $R(\gamma)$ and $R(\Gamma)$ respectively we
denote the range of the graph. 

If the list of possible contractors is properly chosen, then the
states $\{|\Gamma \rangle = |\gamma, j, c \rangle\}$
form an orthonormal basis of ${\cal H}$.  Let us choose once and for all a
particular such orthonormal basis ${\cal B}$.  An important point is that,
for $|\gamma,j,c\rangle \in {\cal B}$, the set of allowed contractors $c$
is finite for a fixed pair $(\gamma,j)$.  This means that any
spin network is a finite linear combination of states in ${\cal B}$. It
follows that the space $\T$ of finite linear combinations of spin networks
is also the space of finite linear combinations of states in ${\cal B}$.

In order to remove the regulators, \cite{I} required the constraints
to act on `diffeomorphism-invariant' states.  While no state in
${\cal H}$ is invariant under all diffeomorphisms, 
a space of diffeomorphism invariant states was constructed in \cite{ALMMT}.
This was done by working in a larger space which consists of linear 
functionals
on some dense subspace of ${\cal H}$.
We will take this dense subspace to be $\T$ and consider the space
of all linear functionals on $\T$,  
the dual $\T'$ of $\T$.  Because the elements of $\T'$ are
linear functions on $\T$, they will be denoted by `bra' vectors $\langle
\psi | \in \T'$.  Note that if one chooses the topology on $\T$ to be just
that due to its linear structure, the algebraic and topological duals of
$\T$ coincide.   Our spaces satisfy the
relation
\be
\T' \supset {\cal H} \supset \T
\ee
and are analogous to a rigged Hilbert triple.  Since
smooth diffeomorphisms of $\Sigma$ act on $\T$, they have a natural (dual)
action on $\T'$.  The space
$\T'$ is quite large, and in particular contains many linear
functionals which are invariant under the action of all such
diffeomorphisms.  We use $\T'_{Diff}$ to denote the space of such
diffeomorphism invariant functionals.   

It is on this space that the
{\it un}regulated constraints $\hat{H}(N)$ of \cite{I} were 
defined\footnote{For the reader familiar with \cite{I,II,III}
we should caution that these constraints
were denoted $\hat{H}'(N)$ in those works.  In order to reduce
the already formidable amount of notation present in this paper, 
we will not explicitly differentiate between the action of an operator
on a space $\T$ and the dual action of the operator on the space of linear 
functionals on $\T'$.   In addition, we will explicitly display the
regulators for regulated constraints so that $\hat{H}(N)$, with no
regulator, will always denote an operator that acts in the dual space.}.
However, 
because a given constraint $\hat{H}(N)$ depends on a choice of lapse 
function $N$, the constraints themselves are not diffeomorphism
invariant.  Thus, the action of $\hat{H}(N)$ in general yields
a state that is {\it not} diffeomorphism invariant.  Products
such as $\hat{H}(N)\hat{H}(M)$ are therefore not, a priori, defined.
If one wishes to compute commutators, one
needs a space larger than $\T'_{Diff}$ in which to work.

Because the constraints of gravity enforce diffeomorphism
invariance, $\T'_{Diff}$ may be expected to contain any `physical'
states (in the sense of Dirac\cite{Dirac}).
However, in the current work we are interested in
the constraint algebra, which must vanish on physical states.
In fact, the classical commutator \cite{BD,TR} of two Hamiltonian
constraints becomes trivial
when just the diffeomorphism constraint is satisfied, so we
again see that
$\T'_{Diff}$ is too small for our purposes.  We now introduce
a larger space $\T'_*$ of `vertex-smooth' states with  $\T'_{Diff}
\subset \T'_* \subset \T'$.  

\subsection{The vertex-smooth states}
\label{vss}

We seek a space which carries a well-defined action of
the constraints of \cite{I} and which is
preserved by that action.  The fact \cite{I}
that the constraints are `anomaly-free' (in the sense defined in \cite{I})
on diffeomorphism invariant states
may be taken as a hint that such a space should exist.
Furthermore, we would like the natural action of the
diffeomorphism group to give a faithful representation on this new space.
That is to
say, only the identity diffeomorphism should be represented trivially.

Readers who are already familiar with the constraints introduced
in \cite{I} will recall that those constraints were defined only
on diffeomorphism invariant states.  Specifically it was important
that the action of the (dual) state
$\langle \psi | 
\in {\cal T}'_{Diff}$
on a spin network over a graph
$\gamma$ does
not depend on the exact placement of the edges of $\gamma$.  This is true
for any diffeomorphism invariant state, 
as its action remains the same when an edge is moved
by a small diffeomorphism.  The key point concerning our new space is
that its states, too, will not care about the exact placement of edges, 
yet they will care about the placement of {\it vertices}
As a result, 
the space ${\cal T}'_*$ will carry a faithful representation of the
diffeomorphism group. The
careful reader may object that moving an edge generally involves
moving vertices as well, but this will be dealt with in section III.

Our definition is as follows.  Let $\C_*' \subset \C'$ contain those
$\langle \psi |$ such that:

\begin{itemize}
\item{A)} if two spin networks $\Gamma_1$ and $\Gamma_2$ are related by a smooth
diffeomorphism which is the identity on their vertices, then 
\be 
\langle \psi | \Gamma_1 \rangle = \langle \psi | \Gamma_2 \rangle.
\ee

Thus, if we fix some `reference' spin network $\Gamma^0$ with $k$ vertices
$v_1, ....,v_k$, then, as $\varphi$ ranges over $Diff^\infty(\Sigma)$, 
$\langle \psi | {\D}_\varphi
|\Gamma^0 \rangle$ is some function 
of the $k$-tuple $(\varphi(v_1),...,\varphi(v_k))$ of vertices of 
$\Gamma^0 \circ \varphi$.  That is to say that $\langle \psi | {\D}_\varphi
| \Gamma^0 \rangle$ is described by a function $\tilde \psi_{\Gamma_0}$
on the space of maps $ \varphi|_{V(\Gamma)}: V(\Gamma) \rightarrow \Sigma $ 
given by restricting diffeomorphisms $\varphi$ to $V(\Gamma)$.  

\item{B)} Each function $\tilde\psi_{\Gamma^0}$ as above extends to a {\it 
smooth}\footnote{For many purposes,
it would in fact be sufficient
to use continuous functions $f$.  However, requiring $\psi_{\Gamma_0}$
 to be differentiable
allows infinitesimal diffeomorphisms to act on $\T'_*$, and taking 
$\psi_{\Gamma_0}$ to be smooth allows $\T'_*$
 to be preserved under this action.}
function $\psi_{\Gamma^0}: \Sigma^{V(\Gamma)}
\rightarrow {\bf C}$ on the {\it entire} space $\Sigma^{V(\Gamma)}$
of maps $\{\sigma :
V(\Gamma) \rightarrow \Sigma \}$ from $V(\Gamma)$ to $\Sigma$.  
In particular, 
$\psi_{\Gamma^0}$ must be smooth at points where
two or more vertices are mapped to the same point in $\Sigma$, 
despite the fact that $\tilde{\psi}_{\Gamma_0}$
was only
defined on maps $\sigma$ that take distinct vertices to distinct points.

\end{itemize}

As a result, a state $\langle \psi | \in
\T_*'$ can be characterized by a family of
smooth functions $\psi_\Gamma:\Sigma^{V(\Gamma)} \rightarrow {\bf C}$, one
for each equivalence class of spin networks under smooth diffeomorphisms.
The diffeomorphism invariant elements of $\T'$ are just those states
$\langle \psi|$ for which each $\psi_\Gamma$ is a constant function. 
Thus, $\C_*' \supset \C_{Diff}'$.

We will see below that the Euclidean
constraints of \cite{I,II,DPR,Lee} are well-defined
on this space and that they map this space into itself, allowing us
to compute their algebra.

\section{RST-like operators and their commutator}
\label{RST}

In this section we discuss a general class of operator families which
we call the
the `Rovelli-Smolin-Thiemann-like' operators or the `RST-like' operators.
Such a family is labeled by a lapse function $N : \Sigma \rightarrow {\bf C}$,
as are the Hamiltonian constraints of gravity.  This class will include
the (so-called `nonsymmetric') 
constraints introduced in \cite{I}. 
We show below that all such operators
are defined on $\T'_*$ and map $\T'_*$ into itself.  We will also show
that any two operators $\hat{H}(N)$ and $\hat{H}(M)$ in the 
same family commute.

\subsection{The Regulated operators}

RST-like operators are based on the notion 
of a `loop assignment scheme' $\alpha$, which takes a vertex $v$
of a graph $\gamma$ and an ordered
pair $(I,J)$ of edges in $\gamma$ and assigns to $(\gamma,v,I,J)$
a smooth loop $\alpha(\gamma,v,I,J) : [0,1] \rightarrow \Sigma$.  
Below, we use the symbol $\alpha(\gamma,v,I,J)$ to denote either
the map from $[0,1]$ to $\Sigma$ or 
its orientation preserving reparametrization  invariance class; the 
meaning should  be clear from  the context. For the purposes of this
paper, we require a loop assignment scheme to have the following properties.
1) Each loop $\alpha(\gamma,v,I,J)$ must begin and end at $v$ and
be such that  $R(\g)\cup R(\alpha(\gamma,v,I,J))$
is a subset of the range $R(\g')$ some other
graph $\g'$; for example, this excludes loops with infinitely many
self-intersections.
2) The loop $\alpha(\gamma,v,I,J)$ is also required to
span a nontrivial area and to be tangent to the plane
defined by $(I,J)$ at its beginning and its end.
3) Finally, a loop assignment scheme must be
`locally diffeomorphism covariant,' in the sense that if $(\gamma,v,I,J)$
restricted to a neighborhood  $W$ of $v$ 
is related to $(\gamma',v',I',J')$ restricted to a neighborhood
$W'$ of $v'$  by a smooth diffeomorphism $\varphi$, then
$\alpha(\gamma',v',I',J') = \varphi' \circ \alpha(\gamma,v,I,J)$ where
$\varphi'$ is some (possibly different) smooth diffeomorphism
which coincides with $\varphi$ on the restriction of
$(\g,v,I,J)$ to $W$. Here, the 
symbol $\circ$ denotes the composition of functions.

The loop assignment scheme will
play the role of a `regulator' for the quantum operator with the idea
that, as the regulator is removed, one should pass through a series of
loop assignment schemes in which the loops shrink to points.  This
limit will be discussed in more detail shortly.

Having chosen a loop assignment scheme $\alpha$, a
{\it regulated} RST-like operator is constructed from a family 
of operators 
\be \label{HT}
\h^\alpha(x): \T \rightarrow \T
\ee
associated with the points $x\in \Sigma$.
The action of $\hat{h}^\alpha(x)$ on
a spin network $|\Gamma \rangle = |\g,j,c\rangle$ 
vanishes when $x$ is not a vertex of $\g$, and otherwise can be 
written in the form
\be\label{RSTlike}
\h^\alpha(x)|\gamma,j,c\rangle\ =\ \sum_{I,J \in E(\gamma,v)}
U^i[\alpha(\gamma,x,I,J)] 
|\g,j,h_i(\gamma,j,x,I,J)c\rangle,
\ee 
where $I,J$ are members of the set $E(\gamma,v)$ of edges of $\gamma$
incident at $v$
and there is a vector of linear operators
\be
h_i(\gamma,j,x,I,J):c\ \mapsto\ h_i(\gamma,j,x,I,J)c,   
\ee
on the space of contractors for $\gamma,j$
associated to every pair 
of edges $(I,J)$ intersecting at the point $x$.
  The repeated index $i$ is summed over $i\in {1,2,3}$
and $U^i[\alpha]$ is the traceless part of the holonomy $U[\alpha]$
defined by
\be
U[\alpha] = U^0[\alpha]\openone + U^i[\alpha]\tau_i,
\ee
where $\tau_i$ are the generators of $SU(2)$.
The operator $h_i(x,I,J)$ 
transforms according to the 
the adjoint representation of $SU(2)$ under 
gauge transformations at $x$, is
antisymmetric in $(I,J)$, and
carries a gauge invariant intertwinor $c$  into a vector of 
intertwinors $h_i(x,I,J)c$ 
by changing only the linear operators assigned by $c$ to the particular
vertex $x$.  These operators must
again satisfy a `local diffeomorphism covariance' condition in the sense
that if $(\gamma,j,x,I,J)$ restricted to  $W$ is related to 
$(\gamma',j',x',I',J')$ restricted to $W'$ by
a diffeomorphism $\varphi \in Diff(\Sigma)$ with $\varphi(W) = W'$
for open sets $W \ni x$, $W' \ni x'$, then
$h_i(\gamma,j,x,I,J)$ and $h_i(\gamma',j',x',I',J')$ 
are related by the same diffeomorphism; 
specifically, 
\be
{\D}_\varphi \left( U^i[\alpha(\gamma,x,I,J)] | 
\gamma,j,h_i(\gamma,j,x,I,J)c \rangle \right)
=  U^i[\varphi \circ \alpha (\gamma,x,I,J) ]
| \gamma',j',h_i(\gamma',j',x',I',J')c \rangle.
\ee

Given a loop assignment scheme $\alpha$ and a smooth lapse function $N$, 
the regulated constraint $\hat{H}^\alpha(N)$ is defined by:
\be 
\label{addlapse}
\hat H^\alpha(N) = \sum_{x \in \Sigma} N(x) \hat{h}^\alpha(x).
\ee
The (uncountably infinite) sum is well defined when acting
on an element $|\phi \rangle$ of $\T$ as 
all but a finite number of terms annihilate any given such $|\phi\rangle$. 

It is clear from \cite{I} that the regulated `non-symmetric'
constraints proposed in that work are of the form (\ref{RSTlike})
and define regulated RST-like operators.  The same is true of the
constraints discussed in \cite{DPR} (which are related to those of
\cite{I} by `changing the factor ordering').
For these particular proposals, the loop assigned to any vertex $v$
and edge pair $(I,J)$ first runs along $I$, then crosses over to $J$
without intersecting any other edges, and returns to $v$ along $J$.
The details of $h_i(\gamma,j,x,I,J)$ for the proposal of \cite{I} 
depend on the choice of
volume operator and on the particular interpretation of the regularization
scheme\footnote{Di Pietri has pointed out \cite{DiP}
that the construction given in
\cite{I} explicitly excludes the possibility of the constraints acting at
planar vertices, due to its reliance on (nondegenerate) tetrahedra.  This
limitation is easily removed, and our discussion includes both cases, 
with either the volume operator of \cite{RSarea} or that of \cite{ALvolume}.}.

\subsection{Removing the regulator}
\label{takelim}

Having defined the regulated operators, the regulator $\alpha$ is now
to be `removed' by considering sequences $\{\alpha_n : n \in {\bf Z}, n \ge 0\}$
in which, as $n 
\rightarrow \infty$, the loops $\alpha_n(\gamma,v,I,J)$ shrink to
the vertex $v$, and such that loops $\alpha_n(\gamma,v,I,J)$
which correspond to the same graph, vertex, and edges but to
different values of $n$ are related by
diffeomorphisms which map the graph $\gamma$ to itself.  That is to
say that $\alpha_n$ should satisfy $\alpha_n(\gamma,v,I,J)
= \varphi_n \circ \alpha_0(\Gamma,v,I,J)$ for some $\varphi_n 
\in Diff(\Sigma)$ such that $\varphi_n$ preserves the edges of 
$\gamma$ (and their orientations) and $\varphi_n(v)
=v$ for all $v \in V(\gamma)$.  
The sequence should also be such that, given $(\gamma,v,I,J)$ and 
an open set
$W \ni v$, there is some $\tilde n$ for which, for all $n \ge \tilde n$, 
we have $\alpha_n(\gamma,v,I,J) \subset W$ and $\varphi_n \circ
\varphi_{\tilde n}^{-1}$ is the identity outside of $W$.
The `unregulated' constraint operator is to be defined through
\be
\label{op}
\hat{H}(N)|\psi\rangle = \lim_{n \rightarrow \infty} \hat{H}^{\alpha_n}(N)
|\psi\rangle.
\ee
We may schematically denote
this limit by $\hat{H}(N) = \lim_{\alpha \rightarrow 0} \hat{H}^\alpha(N)$,
though the final object $\hat{H}$ will  depend on
the particular sequence of loop assignment schemes chosen.  Such 
an object $\hat{H}(N)$ will be called an (unregulated) RST-like 
operator.

Note that, when acting on ${\cal H}$, this limit does not
converge at all: typically, for a spin network $|\Gamma\rangle$, 
$H^{\alpha_n}|\Gamma\rangle$ is orthogonal to $H^{\alpha_m}|\Gamma \rangle$
for $n \neq m$ because the two states are supported on graphs occupying
different positions in $\Sigma$.
It is interesting to note, however, that (as remarked in \cite{I})
if cylindrical
functions are viewed as functions on {\it continuous} (i.e., nondistributional)
connections, then the limit (\ref{op}) does converge when acting 
on such functions, but the result is just
the zero operator.  This follows from the fact that, as the loops
shrink to a point, the holonomies $U[\alpha_n(\gamma,v,I,J)]$ become
$\openone$ so that $U^i[\alpha_n]$ goes to zero.
Nonetheless, a well-defined non-zero 
limit will be obtained by considering the
dual operator induced by $\hH^{\alpha}$ in the space 
$\T'_*$ of linear 
functionals defined above.

To show this, let us consider $\langle \psi|\hat{H}^{\alpha_n}(N)|\Gamma \rangle$
and take $|\Gamma\rangle = |\gamma,j,c\rangle$ to be a spin network.
For each $n$, the functions
$U^i[\alpha_n(\gamma,v,I,J)]|\gamma,j,h^i(\gamma,j,x,I,J)
c\rangle$ are all cylindrical
over some graph $\gamma_n$, and $\gamma_n$ can be chosen such
that $\gamma_n = \varphi_n(\gamma_0)$  where $\{ \varphi_n \}$
is the sequence of diffeomorphisms described in the definition of an
RST-like operator above.

We would like to decompose the function
$U^i[\alpha_n(\gamma,v,I,J)]|\gamma,j,h^i(\gamma,j,x,I,J)c\rangle$
as a sum of spin
networks in our basis ${\cal B}$.
  The important point is that only a finite number of spin
network states can appear in this decomposition.  This is because the
allowed spin is bounded by the sum of the maximum spin in the list $j$ and
($1/2$ times) the maximum number of times the loop $\alpha_n(\gamma,v,I,J)$
retraces itself.  As a result, we may write
\be
\label{expand}
\sum_{I,J \in  E(\gamma,v) }  U^i[\alpha_n(\gamma,v,I,J)] |\gamma,j,
h^i(\gamma,j,x,I,J)c \rangle
= \sum_{k=1}^{K^v} a^k_{v,n} |\Gamma_{v,n}^k \rangle
\ee
for some $K^v \in {\bf Z}$, where we have explicitly indicated
that the coefficients $a^k_{v,n}$, the spin networks $\Gamma_{v,n}^k$, and
the integer $K^v$ will depend on the vertex $v$.   Note that the
$|\Gamma^k_{v,n}\rangle$ can be chosen so that $|\Gamma^k_{v,n}\rangle
= {\D}_{\varphi_n} |\Gamma^k_{v,0} \rangle$, in which case the local
diffeomorphism covariance of the loop assignment guarantees that the
coefficients $a^k_{v,n}$ are in fact independent of $n$.  We will assume
that such a choice has been made and write $a^k_{v} := a^k_{v,n}$. It 
then follows that the action of $\langle \psi | \in \T'_*$ on 
such a state is 
\be
\label{prelim}
\sum_{I,J \in E(\gamma,v)} \langle \psi |
U^i[\alpha_n(\gamma,v,I,J)] | \gamma,j,h_i(\gamma,v,I,J)c\rangle 
= \sum_{k=1}^{K^v} a_v^k \psi_{\Gamma_{v,0}^k} ( \varphi_n|_{V(\Gamma^k_{v,0})}
), \ee
where $\varphi_n|_{V(\Gamma^k_{v,0})} : V(\Gamma^k_{v,0}) \rightarrow 
\Sigma$ is just the map obtained by restricting $\varphi_n$ to
$V(\Gamma^k_{v,0})$.

Taking the limit $n \rightarrow \infty$ amounts
to simply moving around the vertices $V(\Gamma^k_{v,0})$.
We note that a vertex
$v' \in V(\Gamma^k_{v,0})$ is either a vertex of the original graph $\gamma$
(in which case it is mapped to itself by $\varphi_n$), or a point
on one of the curves $\alpha_0(\gamma,v,I,J)$.  Since the sequence
$\{\varphi_n \}$ contracts all points of $\alpha_0(\gamma,v,I,J)$ to $v$ and
the collection of such curves is finite, the limit of (\ref{prelim})
as $n\rightarrow \infty$ is given by replacing $\varphi_n$ on the right-hand
side of (\ref{prelim}) with $\varphi_\infty$, where
$\varphi_{\infty}(v') = \lim_{n \rightarrow \infty} \varphi_n (v')$ is
well-defined for $v' \in V(\Gamma^k_{v,0})$. Thus,
$\langle \psi| \hat{H}(N) = \lim_{n \rightarrow \infty} \langle \psi|
H^{\alpha_n}(N)$ is a well-defined element of $\T'$.  

Note that in fact, for $v' \in V(\Gamma^k_{v,0})$, $\varphi_\infty(v')$ 
is a vertex of the graph $\Gamma$; this reminds us of the implicit 
dependence of $\varphi_n$ on $V(\Gamma)$.  
Due to the local diffeomorphism covariance of the loop assignments, 
(\ref{lim}) is clearly unchanged when the spin-network $\Gamma$ is
replaced by $\phi (\Gamma)$ for $\phi \in Diff^{\infty} (\Sigma)$
such that $\phi$ is the identity on $V(\Gamma)$.  For a $\phi$
that does act nontrivially on $V(\Gamma)$, it changes only the 
points to which $\varphi_\infty$ contracts the vertices of $V(\Gamma^k_{v,0})$.
As a result, we may rewrite the $n \rightarrow \infty$ limit of
(\ref{prelim}) by introducing
a map
$\eta^{k,v}: \Sigma^{V(\Gamma)} \rightarrow 
\Sigma^{V(\Gamma^k_{v,0})}$ which takes an assignment $\sigma: V(\Gamma)
\rightarrow \Sigma$  of points in $\Sigma$ to vertices of $\Gamma$
and generates a new assignment 
$\eta^{k,v} (\sigma) : V(\Gamma_{v,0}^k) \rightarrow
\Sigma $ of points $\Sigma$ to vertices of $\Gamma^k_{v,0}$.  The new 
assignment will in general send many vertices to the same point of 
$\Sigma$ and is given by $[\eta^{v,k} (\sigma)] (v') = 
\sigma (\varphi_\infty (v'))$. Thus, 
\be
\label{lim}
\lim_{n \rightarrow \infty} \langle \psi | H^{\alpha_n}(N) {\cal D}_\phi
| \Gamma \rangle =
\sum_{v \in V(\Gamma)} N(v) \sum_{k = 1}^{K^v} a^k_v \psi_{\Gamma_{v,0}^k}
\circ \eta^{k,v} (\phi |_{V(\Gamma)})
\ee
for any $\phi \in Diff(\Sigma)$.  Finally, 
each $\psi_{\Gamma^k_{v,0}} \circ \eta^{k,v}$ depends smoothly
on the map $\phi|_{V(\Gamma)}$, and extends to a smooth function on all of
$\Sigma^{V(\Gamma)}$..  As a result, 
$\langle \psi |\hat{H}(N)$ is an element of $\T'_*$ and 
$\hat{H}(N)$ maps $\C'_*$ into itself, as desired.
Note that,
because of the covariance of the loop assignments and the operators
$h_i(\gamma,j,x,I,J)$,  
the operator $\hat H (N)$ is also covariant in the sense that
\be
{\D}_\varphi \hat{H}(N) {\D}^{-1}_\varphi = \hat{H}(N \circ \varphi).
\ee

It is clear that the space $\T'_*$ could in fact be extended even
further.  Just as it was sufficient for our states to depend 
smoothly on the positions of the vertices, it is not necessary for the 
states to be completely independent of the placement of the edges.
Thus, one could replace requirement A in the definition of $\T'_*$ with
a condition more like that of B,
requiring that the states depend sufficiently
smoothly on the positions of edges so that the limit defining the
unregulated RST-like operators can still be taken.  However, the space
$\T'_*$ defined in section \ref{vss} is enough for our purposes and we
will not discuss further generalizations here.

\subsection{The commutator.}
\label{comsec}

We are now in a position to study the commutator $[\hat{H}(N),
\hat{H}(M)]$ as an operator on $\T'_*$.  Although we work on
a larger space $\T'_*$ and with the unregulated operators (which are
of a more general form than in \cite{I}), the
following argument is much like the anomaly-free calculation of
\cite{I}.  We will proceed by choosing some
$\langle \psi | \in \C'_*$  and some spin network $\Gamma$.  We wish
to evaluate $\langle \psi | [\hat{H}(N),\hat{H}(M)] |\Gamma \rangle$
for all $N,M \in C^\infty(\Sigma)$.  It will be easiest to compute
the answer in the special case where $N$ vanishes at all vertices of
$\Gamma$ except $v_N$, and $M$ vanishes at all vertices of $\Gamma$
except $v_M$.  The general case can then be reconstructed using the
fact that $\hat{H}(N)$ is linear in $N$.

It is clear from (\ref{lim}) that we may write
\be
\label{two}
\langle \psi | \hat{H}(N) \hat{H}(M) 
|\Gamma \rangle = N(v_N)M(v_M)
\psi^{v_N,v_M}
\ee
for some complicated function $\psi^{v_N,v_M}$. Note that
when $v_N = v_M$, the right hand side is symmetric in $N$ and $M$.
As a result, for this case we have $\langle \psi | [\hat{H}(N),
\hat{H}(M)]|\Gamma\rangle = 0$.   

Let us therefore consider $v_N \neq v_M$.  
The case where $N(v_N)=0$ or
$M(v_M) = 0$ is trivial, so we will assume $N(v_N) \neq 0$
and $M(v_M) \neq 0$.  Since
(\ref{two}) depends on the values of $N$ and $M$ only at $v_N$ and $v_M$
respectively, the result (\ref{two}) is unchanged if $N,M$
are replaced by smooth functions $\tilde N, \tilde M$,
such that $\tilde N (v) = N(v)$, 
$\tilde M (v) = M(v)$ for all $v \in V(\Gamma)$ but for which
the support (${\rm supp} \tilde N$) of $\tilde N$ does not intersect the
support (${\rm supp} \tilde M$) of $\tilde M$.

The action of $\hat{H}^{\alpha_m}(\tilde M)$ on $|\Gamma\rangle$ is, 
\be
\label{first}
\hat H^{\alpha_m}(\tilde M) |\Gamma \rangle = \tilde M(v_M)
\sum_{I,J \in E(\gamma,v)}
U^i[\alpha_m(\gamma,v_M,I,J)]|\gamma,j,h_i(\gamma,j,v_M,I,J) c \rangle.
\ee
Since the support of $\tilde M$ is an open set containing $v_M$,
for $m$ greater than or equal to some
$\tilde m$ we must have  $R(\alpha_m(v_M,\gamma,I,J)) \subset {\rm supp}
\tilde M$ for all $I,J$.  Choosing $m \ge \tilde m$, let us now act on
(\ref{first}) with $\hat H^{\alpha_n}(\tilde N)$.  Note that because
$R(\alpha_m(\gamma,v,I,J)) \cap {\rm supp} \tilde N$ is empty,
$\hat{H}^{\alpha_n}(\tilde N)$ will act only at vertices of the
original graph $\gamma$.\footnote{That easy observation is  crucial here; 
the reader familiar with \cite{I} knows that before, it was thought that 
an action of the second Hamiltonian
on the vertices produced by the first one was relevant for the result.}  
In particular, it will act only at $v_N$.  
Thus, it can be shown that for sufficiently large $n$ and $m$, we 
have

\ba 
\hat H^{\alpha_n}(\tilde N) \hat{H}^{\alpha_m}(\tilde M) |\Gamma \rangle
\sim
\tilde N(v_N) \tilde M(v_M)  \nonumber\\ \sum_{{{I_1,J_1 \in E(\gamma,v_N)}
\atop {I_2,J_2 \in E(\gamma,v_M)}}} 
U^i[\alpha_{n}(\gamma,v_N,I_1,J_1)] U^j[\alpha_m(\gamma,v_M,I_2,J_2)]
 \nonumber\\ \times
|\gamma,j,h_i(\gamma,j,v_N,I_1,J_1)h_j(\gamma,j,v_M,I_2,J_2) c \rangle
\ea 
where $\sim$ denotes equality modulo 
terms which are annihilated by any $\langle \psi | \in \C'_*$ 
in the limit $n,m\rightarrow \infty$. 
Indeed, due to local diffeomorphism covariance of the  linear 
operators $h_i(\gamma, j, v, I,J)$, the operator
$h_i(\gamma,j,v_N,I_1,J_1)$ above acts in the same way, independent of
whether
we first attach the loop $\alpha_m(\gamma,v_M,I_2,J_2)$ or not.
For the analogous reason, the  loop  $\alpha_{n}(\gamma,v_N,I_1,J_1)$ 
is independent, modulo appropriate diffeomorphisms, of 
whether we first attach the loop $\alpha_m(\gamma,v_M,I_2,J_2)$ or not.
The rest is assured by properties A and B of the vertex-smooth states
$\langle \psi | \in \C'_*$. 

Since the above result is symmetric with respect to $(N,M)\rightarrow (M,N)$
we find 
\be
\label{zero}
\langle \psi | [\hat{H}(\tilde N) , \hat{H}(\tilde M) ] |\Gamma \rangle =0.
\ee
{}From our general considerations at the beginning of this subsection, since
(\ref{zero}) holds for arbitrary $\langle \psi |$ and $| \Gamma \rangle$, the
commutator of $\hat{H}(N)$ and $\hat{H}(M)$ must vanish
identically for all smooth $N$ and $M$.  We again
stress that this holds for any RST-like operator, whether it
acts on planar vertices or not. This is our main result.

\section{Symmetrized Operators}
\label{sym}

Let us now turn to the issue of `symmetrizing' the Hamiltonian operators, 
in the sense of adding some kind of `hermitian conjugate
operator' $\hat H^\dagger(N)$ to $\hat H(N)$.
Note that we have not defined an inner product on $\C'_*$; indeed, appendix
A shows that a fully satisfactory such inner product does not exist either on
$\T'_*$ or on any subspace that both A) contains at least one diffeomorphism
invariant state $\langle \psi |_{Diff}$  for which
$\langle \psi|_{Diff} \hat{H}(N) \neq 0$ (for some $N$) and B)
is preserved by the action of a family of Hamiltonian constraints. 
Thus, the operators $\hat{H}(N)$ do not act in a Hilbert
space and there is no canonical notion of whether $\hat{H}(N)$
is `symmetric' or of how to make it so.
However, for the special
case of constant lapse, $\hat{H}(1)$ is invariant under
diffeomorphisms and maps diffeomorphism invariant states into diffeomorphism
invariant states ($\hat{H}(1): \T'_{Diff} \rightarrow \T'_{Diff}$).
As described in \cite{ALMMT}, a family of natural Hermitian inner 
products can be introduced on a subspace (which we shall call 
$\tilde \T'_{Diff}$) of $\T'_{Diff}$, and this subspace may then
be completed to a Hilbert space ${\cal H}_{Diff}$.  The precise
Hilbert space obtained depends on which member of the family of hermitian
inner products was chosen, but we will not indicate this dependence
explicitly.  We note that a particular choice was made in \cite{III}, but
we will instead leave the relevant parameters arbitrary.
For a generic loop assignment (such that the attached loop
$\a(\g,v,I,J)$ never overlaps $\g$),  $\hat{H}(1)$
is a densely defined operator in  ${\cal H}_{Diff}$.
For other loop assignments, $\hat{H}$ may only be defined on some
smaller domain (say, $\Phi$) as states in ${\cal H}_{Diff} \setminus
\Phi$ are carried out of
${\cal H}_{Diff}$ by $\hat{H}(1)$.   In any case, one
can see if $\hat{H}(1)$
is symmetric in the sense of a bilinear form on its domain.
Because $\hat{H}(1)$ typically `destroys'
edges of graphs and usually does not `create' edges, $\hat{H}(1)$ is 
not symmetric in any of the Hilbert spaces ${\cal H}_{Diff}$ unless
the inner product is chosen so degenerate that $\hat{H}(1)$
is just the zero operator in ${\cal H}_{Diff}$.  In this sense then, 
no family of RST-like constraints is symmetric.
                  
The view has been expressed \cite{I,II,DPR,RR} that one might like to
have a family of Hamiltonian constraints that {\it are} symmetric
in some sense. A minimal requirement might be that
$\hat{H}(1)$ defines
a symmetric operator on some natural domain in 
${\cal H}_{Diff}$.  The status of this
view is not completely clear, as there are general arguments
\cite{KK}  that, due to the structure of the classical constraint
algebra, the Hamiltonian constraints should not be self-adjoint.
Since, however, the classical commutator of Hamiltonian constraints
vanishes on the
surface in phase space satisfying the diffeomorphism constraints, this
argument need not apply to $\hat{H}(1)$ on ${\cal H}_{Diff}$. 
See also the commentary of \cite{I} on this issue.  We therefore
wish to consider a `symmetrization' of our family of constraints which
gives a self-adjoint $\hat{H}(1)$ on some ${\cal H}_{Diff}$ and compute
the algebra of the resulting constraints. We will refer to a family
of constraints which satisfies this property as being `constant lapse
symmetric.'

\subsection{Review of Symmetrization Proposals}

The set of proposals \cite{II,DPR,RR} for `symmetrizing' the constraints is
in fact quite diverse.  In this subsection, 
we quickly review the proposals for unregulated symmetric
constraints which have appeared in the literature
before discussing a new (and, we believe,
more satisfying) definition of `symmetrization'
in section \ref{symalg}.  However, we then show (in section IV C) that
even this new definition of symmetrization leads to either
commuting constraints or to constraints which are anomalous in the sense that
their commutator does not even vanish on diffeomorphism
invariant states.  It is in this sense that we use the terms
`anomaly' or `anomalous' in the rest of this work.

At first, it may seem natural to use the Hilbert space structure of
${\cal H}$ to symmetrize the regulated constraints $H^{\alpha_n}(N)$
before taking the limit $n \rightarrow \infty$.  This possibility was
mentioned in \cite{I} and was used as a basis for \cite{RR}.  
After all, for appropriate loop assignment schemes, one can
arrange for a given spin network $|\Gamma' \rangle$ to
appear on the right hand side of the decomposition (\ref{expand}) for
only a finite number of spin networks $|\Gamma \rangle$ and, with this 
arrangement,  
it is true that $[H^{\alpha_n}(N)]^\dagger$ maps $\T$ to $\T$ and so
has a dual action $\T' \rightarrow \T'$.
Unfortunately, 
this method of `symmetrization' fails to define a constant lapse
symmetric family of constraints.

  To see this, consider,
for example, $\langle \psi | [H^{\alpha_n}(N)]^\dagger |
\Gamma \rangle  = \sum_{\Gamma'} \langle \psi |\Gamma'\rangle
\overline{ \langle \Gamma | \hat{H}^{\alpha_n}(N)| \Gamma' \rangle }$
for $\langle \psi | \in \T'_*$ and $|\Gamma \rangle 
= |\gamma,j,c \rangle$, where the sum is over an orthonormal basis of
spin networks $\Gamma'$ and the matrix elements $\langle \Gamma | 
\hat{H}^{\alpha_n}(N) | \Gamma' \rangle$ are taken in ${\cal H}$.
Recall that $\hat{H}^{\alpha_n}(N)$ basically
adds edges (due to the $U[\alpha_n]$ factor) to $\Gamma'$.  Thus, 
$\langle \Gamma |\hat H^{\alpha_n}(N)|\gamma' ,j',c', \rangle$ typically 
vanishes unless $\gamma'$ is a subgraph of $\gamma$ and $\alpha_n(\gamma',
v,I,J)$ supplies exactly the missing edges (for some $(v,I,J)$).
However, the required edges of $\gamma$ lie at a finite separation from the
vertices of $\gamma'$, while the edges added by $U_i[\alpha_n(\gamma',v,I,J)]$
approach $v$ as $n \rightarrow \infty$.  Since there are only a finite
number of subgraphs of $\gamma'$ and one may use the fact that the
$\alpha_n(\gamma', v, I, J)$'s for various $n$ are related by diffeomorphisms
to bound the change in spin, for fixed $|\Gamma \rangle$
there is only a finite set of $\Gamma'$
for which $\langle \Gamma |H^{\alpha_n}|\Gamma' \rangle$ can be nonzero, 
independent of the value of $n$.  As a result, for $n$ greater
than some $\tilde{n}$, $\langle \Gamma | H^{\alpha_n}(N) |\Gamma' \rangle=0$
for all $\Gamma'$.  Thus, adding $\lim_{n \rightarrow \infty}
[\hat{H}^{\alpha_n}(N)]^\dagger$
to $\hat H^{\alpha}(N)$ has, in general, no effect whatsoever\footnote{For
some choices of loop
assignment $\alpha$ and operators $h_i(\gamma,v,I,J)$a, $\hat{H}^\alpha(N)$
can occasionally `destroy' an edge (see \cite{LM}).
In that case, due to the fact that we have not
required the loops to shrink in a uniform ($\gamma$ independent) way,
the limit of $[H^{\alpha_n} (N)]^\dagger$ may be nonzero, or
may not even by defined on all of $\T'_{Diff}$.  Still, $\lim_{n\rightarrow}
[\hat{H}^{\alpha_n}(N)]^\dagger$ annihilates `most' of $\T'_*$, and 
certainly does not lead to symmetry of $\hat{H}(1)$.}.

Another proposal was made by De Pietri, Rovelli, and Borrisov in \cite{DPR}
and corresponds to
`changing the factor ordering' of the regulated constraints
of \cite{I}.
However, in our notation their proposal amounts to simply
using a different set of operators $h_i(\gamma, j, v, I, J)$ than
the original proposal of Thiemann.  In fact, under their proposal, 
the fully `symmetrized' operator is still an RST-like operator.
Thus, such operators are not constant-lapse symmetric in the above sense. 
In addition, the calculation of section \ref{comsec} applies
and the commutator of two such constraints vanishes on $\T'_*$.

Finally, we remark that another kind of symmetrization was considered
in \cite{II}, and involved `marking' various edges.  However, this
method applied only to the regulated constraints.  See we are
interested in the unregulated constraints, we will not discuss this 
proposal here.

\subsection{A new definition of the Hermitian Conjugate}
\label{symalg}

It appears that the sort of `symmetrization' which is desired 
\cite{II,RR,part}
is something that does not involve marking 
special edges and which is somehow closer to the symmetrization of
$\hat{H}(1)$ induced by the inner product on ${\cal H}_{Diff}$.
A step in this direction was suggested to us by Thiemann \cite{prit}
and will be described below together with the resulting definition of
the hermitian conjugates $H^\dagger(N)$.  The idea is to rewrite
the definition of
the Hermitian conjugate of $\hat{H}(1)$ induced by the
inner product on ${\cal H}_{Diff}$ in a suggestive form, which can then
be refined to define a family of operators $\hat H^\dagger(N)$ which we
will refer to as the `hermitian conjugates' of $\hat H(N)$.  It is 
important to note that our definition of the hermitian conjugate will 
make use of special properties of $\hat H(N)$, and will not be applicable to
a general family of operators $\hat{A}(N)$ on $\T'_*$ labeled by
lapse functions.  In particular, it does not directly provide a definition of
the hermitian conjugate of $\hat H^\dagger(N)$, or of the `symmetrized'
operator $\hat H^S(N)  = \hat H(N) + \hat H^\dagger(N)$.  As a result, 
this structure in no ways runs counter to the arguments of \cite{KK}.
On the other hand, the family $H^S(N)$ will be constant lapse symmetric, 
as desired.

Let us begin by reviewing
the inner product defined by \cite{ALMMT} on 
diffeomorphism invariant states.  In fact, \cite{ALMMT} considered
a space of spin network states based on analytic graphs and invariance
only under analytic diffeomorphisms.  What we need here is
an extension to our smooth case.  The construction is similar to
that of \cite{ALMMT}, and in fact simpler \cite{BSII}.  For example
there is no issue of `type I' vs. `type II' graphs (see \cite{ALMMT});
in our smooth case, all graphs may be studied together.  Since the
treatment of our case is direct (given the methods of \cite{ALMMT}), 
we simply state the results below; the reader may consult \cite{BSII}
for details.

We will first need a bit of notation.
Recall that an important notion in \cite{ALMMT} was the
group $GS(\gamma)$ of `graph symmetries' of a graph $\gamma$. 
Roughly speaking, this is the group of all embeddings of 
$\g$ into itself.
We define it as follows: consider the `isotropy' group
${\rm Iso}(\gamma) \subset Diff(\Sigma)$ of diffeomorphisms $\varphi$
that map $\gamma$ to itself.
Also, let ${\rm TA}(\gamma) \subset {\rm Iso}(\gamma)$ (the `trivial
action' subgroup) consist of those $\varphi \in {\rm Iso}(\gamma)$ that map
every {\it edge} of the graph $\gamma$ to itself and preserve every edge's
orientation. The trivial action subgroup is normal in
${\rm Iso}(\gamma)$, and
the graph symmetry group is defined to be the quotient:
${\rm GS}(\gamma) = {\rm Iso}({\rm \gamma})/{\rm TA}(\gamma)$.
Given a graph $\g$, the symmetry group ${\rm GS}(\g)$
acts naturally in the linear space spanned by the spin-networks  over
$\g$.  

Now, for a spin network $\Gamma$ over a graph $\g$, we will define
a linear functional 
$\langle \Gamma, 1| \in \T'_{Diff}$.  Suppose that
 $|\Gamma' \rangle$ is a 
spin network over a graph $\g'$ which is diffeomorphic to $\gamma$ (so
that $\phi_{\g' \g}(\gamma) = \g'$ for some $\phi_{\g' \g} \in Diff(\Sigma)$.
Then,  the action of $\langle \Gamma, 1|$ on $|\Gamma' \rangle$ 
is\footnote{
Such a definition may be  obtained
by `averaging $\langle \Gamma |$ with respect
to the action of the group of diffeomorphisms' \cite{ALMMT}.} 
\be
\langle \Gamma, 1 | \Gamma' \rangle\ :=\  
\sum_{s\in {\rm GS}(\g)}\langle (\phi_{\g'\g}\circ s)(\Gamma)  
|\Gamma'\rangle.  
\ee
When $\g'$ is not diffeomorphic to $\g$, the result is just
zero.
If we let $\tilde{\T}'_{Diff}$ be the space spanned by (finite) linear 
combinations of the $\langle \Gamma, 1 |$, then
a natural family of inner products \cite{ALMMT} on
$\tilde{\T}'_{Diff}$ is given by
\be
\label{dip}
\langle \Gamma, 1 | \Gamma', 1 \rangle \ =\
a_{[\g]}\langle \Gamma, 1 | \Gamma' \rangle
\ee
for any set of positive real
constants $a_{[\g]}$ which may depend on the diffeomorphism
class of $\g$. The Hilbert space ${\cal H}_{Diff}$ is just the 
completion of $\tilde \T'_{Diff}$ in one of the inner products (\ref{dip}).

As a result, 
the hermitian conjugate of 
$\hat{H}(1)$ is defined on an appropriate domain
by $\langle \Gamma, 1|\hat{H}^\dagger(1) 
|\Gamma' \rangle = \overline{ \langle \Gamma', 1 | H(1)| \Gamma \rangle}$
where the overline denotes complex conjugation.
This led  Thiemann to suggest \cite{prit} that a `Hermitian conjugate family'
$\hat H^\dagger(N)$ be defined on diffeomorphism invariant states of the
form $\langle \Gamma, 1|$ by some sort of expression of the form:
\be
\label{symondiff}
\langle \Gamma, 1 | H^\dagger(N) | \Gamma' \rangle 
= \overline {   \langle \Gamma', 1 | H(N) {\D}_{\varphi_{\Gamma, \Gamma'}}
|\Gamma \rangle }
\ee
for an appropriate diffeomorphism $\varphi_{\Gamma,\Gamma'}$ that,
in some sense, moves $\Gamma$ to
the location in $\Sigma$ occupied by $\Gamma'$.  Our task it to 
make this suggestion precise, and in fact we will simultaneously
extend it so that $\hat H^\dagger(N)$ can act on states which are
not necessarily diffeomorphism invariant.  Nevertheless, we take
(\ref{symondiff}) as our moral inspiration and link to the inner 
product (\ref{dip}).

To proceed, we will first need to introduce some new
notation.  For example, for every graph $\g$ we define
the pointed symmetry group ${\rm GS}_*(\gamma)$ by replacing, at every stage
in the  the definition of ${\rm GS}(\g)$, the groups $Iso(\gamma)$ and
$TA(\gamma)$ with their subgroups $Iso_*(\gamma)$ and $TA_*(\gamma)$
of diffeomorphisms which are the identity on
the vertices of $\g$. 
Next,  given any spin network
$\Gamma$ and any map $\sigma: V(\Gamma) \rightarrow \Sigma$,
we define the state $\langle \Gamma_\sigma|$ to be the linear
functional for which
\be
\langle \Gamma_\sigma |  \Gamma' \rangle \ =\ 
\sum_{s\in {\rm GS}_*(\g)} \langle \phi^\sigma_{\g'\g}\circ s(\Gamma) | 
\Gamma'\rangle  
\ee
for $|\Gamma' \rangle$ such that there is a diffeomorphism  
$\phi^\sigma_{\g'\g}$ satisfying
$\phi^\sigma_{\gamma'\gamma}(\gamma) = \gamma'$ and which, 
when restricted
to the vertices of $\gamma$, coincides with $\sigma$.  For other
$|\Gamma' \rangle$, we set $\langle \Gamma_\sigma | \Gamma' \rangle = 0$. 
Note that if $\sigma$ happens to map two distinct vertices of $\Gamma$ to
the same point, then $\langle \Gamma_\sigma|$ is just the zero functional.
The states $\langle \Gamma_\sigma|$
 do not lie in $\T'_*$, because their action does not
depend smoothly on the location of the vertices of $\Gamma'$; it vanishes
unless the vertices of $\Gamma'$ occupy exactly
the positions assigned by  the map
$\sigma$.    However, these states can be used to build a large class of
states in $\T'_*$ which we will label $\langle \Gamma, f|$ with $\Gamma$
a spin network and $f: \Sigma^{V(\Gamma)} \rightarrow {\bf C}$ a smooth
function.  These are the states defined by
\be
\label{fdef}
\langle \Gamma, f | \ : = \sum_{\sigma\in \Sigma^{V(\g)}} f(\sigma) \langle 
\Gamma_\sigma|
\ee
 Recall that such (uncountably infinite) sums are well defined in 
$\T'$ as, when acting on a given spin network state, only a finite
number of the terms contribute. For example, given spin networks 
$\Gamma$ and $\Gamma'$ associated to graphs $\g$ and $\g'$ respectively, 
if there is some $\phi_{\gamma' \gamma} \in Diff(\Sigma)$ such
that $\phi_{\gamma' \gamma} (\gamma) = \gamma'$ then we have
\be
\label{action}
\langle \Gamma, f | \Gamma' \rangle = 
 \sum_{s \in GS(\gamma)} \langle (\phi_{\gamma' \gamma} \circ  s) (\Gamma) 
 | \Gamma' \rangle f((\phi_{\gamma' \gamma}
\circ s) |_{V(\Gamma)}). \ 
\ee
Otherwise, we have $\langle \Gamma, f | \Gamma' \rangle =0$.
It follows from the definition, that every state $\langle \Gamma, f|$
is uniquely determined by the sum 
\be
sum_{s\in GS(\g)}{\bar f}(s_{|_{V(\gamma)}}) |s(\Gamma)\rangle
\ee

Before introducing the new definition of the Hermitian conjugate of
a family of RST-like operators, we will need one more bit of notation.
For   any subset $X \subset V(\Gamma)$ of vertices of a
spin-network $\Gamma$, we define
a linear functional on $\T$ corresponding to the notion of $\langle 
\Gamma |$ `averaged with respect to the diffeomorphisms 
acting trivially on $X$',
\be
\T'\ \ni\ \langle \Gamma |^X\ :=\ \sum_{\sigma_{|_X}={\rm id}}\langle 
\Gamma, \sigma|.                                                        
\ee         
The functional $\langle \Gamma | $ is again not an element of $\T'_*$
but the action of $\hH(N)$ is naturally defined on such states  by 
(\ref{lim}).  
                     
For any RST-like operator $\hat H(N)$, we now define a `conjugate' operator
$\hat H^\dagger(N)$ through the following procedure.  When $\Gamma'$
has at least as many vertices as $\Gamma$, we set
\be
\label{hc}
\langle \Gamma, f | H^\dagger(N) |\Gamma' \rangle 
\ =\  
\sum_{\sigma\in \Sigma^{V(\g)}}f(\sigma) \sum_{X \in V_{|V(\Gamma)|}(\Gamma')}
\biggl(\overline{
\langle \Gamma' |^X  
|\hat H
(N)| \phi_\sigma(\Gamma)\rangle } \biggr),
\ee
where $\phi_\sigma$ is any diffeomorphism of $\Sigma$ that coincides with
$\sigma$ on $V(\Gamma)$ and $V_n(\Gamma')$ is the set of all $n$-element
subsets of $V(\Gamma')$ (so that $V_{|V(\Gamma)|}(\Gamma')$ is
the set of all sets of vertices of $\Gamma'$  
which have the same number of elements as
the set of the vertices of $\Gamma$).  When $\Gamma'$ has
less vertices than $\Gamma$, we set $\langle \Gamma, f | H^\dagger(N)|\Gamma'
\rangle =0$. The result is a well-defined family of operators 
$\hat H^\dagger(N) : \tilde \T'_* \rightarrow \T'$.

Implicit in our definition is the fact that a (regulated) RST-like operator
always adds vertices when acting on a spin network $|\Gamma \rangle$.
Thus, the action of an (unregulated) RST-like operator on the (dual) state
$\langle \Gamma, f|$ is of the form $\sum_{\Gamma'} \langle 
\Gamma',f_{\Gamma'}|$ where each $\Gamma'$ has {\it less} vertices
than $\Gamma$.  As a result, we may expect that $\langle \Gamma, f|
H^\dagger (N) = \sum_{\Gamma'} \langle \Gamma', f' |$ where now
each $\Gamma'$ has {\it more} vertices than $\Gamma$.  It is this idea
which motivates the definition of $\hat{H}^\dagger(N)$ given above.

There are several properties of (\ref{hc}) that we would like to point out.
For example, the reader may
readily verify that the  hermitian conjugate family of operators
(\ref{hc}) is consistent with the
Hilbert space inner product on ${\cal H}_{diff}$: when the domains
are properly chosen, the restriction
of $\hH^\dagger(N)$ to ${\cal H}_{diff}$ for $N=1$
is the hermitian conjugate of the associated restriction of $\hat H(1)$
with respect to the Hilbert product of ${\cal H}_{diff}$.  

In addition, one sees that when $\Gamma'$ has at least as many vertices as 
$\Gamma$, the only nonzero contribution
in (\ref{hc}) comes from terms in which  $\sigma(V(\Gamma)) =X$.  Thus
(\ref{hc}) depends on the values of the lapse $N$ only at the vertices of
$\Gamma'$.
In this sense $\phi_\sigma$ moves $\Gamma$ to the position occupied
by $\Gamma'$.  The details of (\ref{hc}) are complicated due to the desire 
to deal carefully with graphs with symmetries; we recall that the action 
(\ref{action}) of $\langle \Gamma, f|$ also has several terms when 
$GS(\gamma)$ is not empty.  In this sense then, (\ref{hc}) can be seen
as a precise implementation of the ideas of (\ref{symondiff}).
It would also appear that (\ref{hc}) is a precise
implementation of the ideas of \cite{coll}.

Finally we note that, as expected, $H^\dagger(N)$ generally
`adds edges':  when the spins of $\Gamma$ are large enough, 
$\langle \Gamma, f | H^\dagger(N) = \sum_{\Gamma'} \langle \Gamma', f^{\Gamma'}
|$ such that $\gamma$ is diffeomorphic to a subgraph of each $\gamma'$.
The reader may also wish to
investigate other properties of this expression for
himself.

Now, a priori, the action of $H^\dagger(N)$ is only defined on
the vector space (which we shall call $\tilde {\cal T}'_*$)
spanned by the states $\langle \Gamma, f |$.  This space
is in fact sufficient
for our study of the commutator, but we may also attempt to extend 
the definition of $H^\dagger(N)$ to all of
$\T'_*$ using a kind of `super linearity.' The point is that
a general state $\langle \psi | \in \T'_*$ can be written as a certain
sum of states of the form $\langle \Gamma, f|$.
Indeed,   for a state 
$\langle \psi | = \langle \Gamma, f|$  such that 
the graph symmetry group of the underlying graph
$\gamma$ is trivial, the function $f$ is just 
the function
$\psi_\Gamma$ from the definition of $\T'_*$ in section \ref{vss},
while $\psi_{\Gamma'}$ for all $\Gamma'$ not diffeomorphic
to $\Gamma$ is zero.   Thus, a general state $\langle \psi | \in \T'_*$
can be written as
\be
\label{basis}
\langle \psi | = \sum_{\Gamma \in S^*} \langle \Gamma, \tilde
\psi_\Gamma |,
\ee
where the sum is over a set $S^*$ of orthonormal spin networks
such that $\T$ is spanned by states for the form ${\D}_\varphi |\Gamma \rangle$
for $\varphi \in Diff(\Sigma)$ and $\Gamma \in S^*$.  For spin networks
$\Gamma$ over graphs which have no symmetries we have
$\tilde \psi_\Gamma = \psi_\Gamma$, while for graphs with nontrivial
graph symmetry groups $\psi_\Gamma$ and  $\tilde \psi_\Gamma$ are
related by a certain symmetrization procedure induced by (\ref{fdef}).
As a result, we may attempt to extend $\hat H^\dagger(N)$ 
to an operator on $\T'_*$ by defining
\be
\label{slin}
\langle \psi | H^\dagger(N) | \Gamma \rangle = \sum_{\Gamma' \in S^*}
\langle \Gamma', \tilde \psi_{\Gamma'} | H^\dagger(N) | \Gamma \rangle.
\ee
Such a sum will make sense if, for each state $|\Gamma \rangle$,
only a finite
number of terms contribute. This is the case if one uses a loop 
assignment such that none of the loops $\a(\g,v,I,J)$ overlaps $\g$.

However, for the assignment used in \cite{I,II,DPR,CL},
we must distinguish two cases in the work of Thiemann. Due to
the phenomenon of `disappearing edges'\cite{LM}, $H^\dagger(N)$ as defined
by the original constraints $\hat H(N)$ of \cite{I} does not
satisfy the above requirement.  That is, for certain choices of 
$|\Gamma \rangle$, 
an (uncountably) infinite number of terms in the sum (\ref{slin}) may
be nonzero.  Thus, we say that the action of $\hH^\dagger(N)$ on a general
element of $\T'_*$ is divergent. Unfortunately, 
this is true even if we attempt to extend the definition only to the
image $\hat{H}(M)[\tilde \T'_*]$ of $\tilde \T_*'$ under $\hat{H}(M)$.
(Again, because of the disappearing edges phenomenon,
the smaller space $\tilde \T'_*$ is not preserved by $\hH^\dagger(N)$).
Thus, it is not possible to define $\hat H(N) \hat H^\dagger(M)$ 
even on $\tilde{\T}'_*$, or
to compute a commutator on this space.  Furthermore, there is no
subspace on which $\hat H(N)$ and $\hat H^\dagger(N)$ are both defined 
and which
is preserved by both of these operators.  Similarly, for this case, 
$\hat H(1) + \hat H^\dagger(1)$ is only defined on a subspace
$\tilde{\cal H}$
of ${\cal H}_{Diff}$ which is not dense in ${\cal H}_{Diff}$. However, 
there are
subspaces $\T'_{*,n}$ of $\T'$
on which the $n$-fold action of $\hat H(N)$ or 
$\hat H^\dagger(N)$
are defined.  Thus, we might calculate the commutator on $\T'_{*,2}$.
More will be said about this shortly.

The other case considered by Thiemann \cite{I,II} is when certain additional
`projection' operators are introduced to remove the offending terms
generated by $\hat H(N)$.  As a result, we must now consider 
the class of `projected RST-like operators' defined in the same
way as the RST-like operators, but from regulated operators of the form
\be
\label{pRSTlike}
\h^\alpha(x)|\gamma,j,c\rangle\ =\ p_\gamma \sum_{I,J \in e(\gamma,v)}
U^i[\alpha(\gamma,x,I,J)] 
|\g,j,h_i(\gamma,j,x,I,J)c\rangle.
\ee 
Here, all is as with the RST-like operators, except for the addition of
the projection $p_\gamma$.  This $p_\gamma$ is just the projection 
onto the space of all spin networks $|\g', j', c' \rangle$ over graphs
$\gamma'$ such that the graph
$\gamma$ is a subgraph of $\gamma'$.  These projections
clearly satisfy their own version of `diffeomorphism covariance' and, 
as a result, the arguments of sections \ref{takelim} and \ref{comsec}
can be repeated for the projected RST-like operators, showing that
they yield well defined operators $\hat{H}(N) : \T'_* \rightarrow
\T'_*$ and that any two such constraints commute.
As before, we do not comment on the motivations
behind this construction, but simply use it to calculate the commutator of
the symmetrized constraints.

With the projectors in place, the associated Hermitian conjugate 
operators $H^\dagger(N)$ defined by (\ref{hc}) are well-defined on all
of $\T'_*$.  We also note 
that for a state $\langle \psi |$
 in $\T'_{*,2}$, the action of a projected operator on
$\langle \psi |$ is {\it identical} to the action of the corresponding
unprojected operator on $\langle \psi |$.  As a result, our
discussion of the projected case below includes much of the information
about the unprojected commutator
In fact, an argument similar to the one given below also holds for the 
unprojected operator on $\T'_{*,2}$ (and produces similar results),
so long as appropriate care is
taken with the convergence of various `superlinear' expressions.  However, 
we will not explicitly deal with this more subtle case.

\subsection{A commutator, again}

We would now like to compute the commutator of the symmetrized
Hamiltonian constraints $H^S(N) = H(N) + H^\dagger(N)$ on $\T'_*$,
with $H^\dagger(N)$ given by (\ref{hc}).
We restrict ourselves to the particular constraints proposed by
Thiemann, and use the projected form (\ref{pRSTlike}).
As stated above, an argument similar to the one below also applies to
Thiemann's unprojected operators on $\T'_{*,2}$, so long as appropriate
care is taken with superlinear expressions.
In addition, we consider only forms of Thiemann's constraint
which do not act at planar vertices -- it is straightforward
to show than any other case is anomalous in the sense that
$[H^S(N),H^S(M)]$ does not annihilate diffeomorphism invariant states.
We will not review the details of Thiemann's proposal here, 
but we remind the reader of what for us is the most important
property of this proposal:  that for a spin network $|\Gamma\rangle$, 
the graphs $\gamma^k_{v,0}$ associated with
the spin networks $|\Gamma^k_{v,0} \rangle$ appearing in (\ref{expand})
differ from $\gamma$ only in having a single extra edge
attached to two edges that intersect at $v$ as shown below:
{\vskip .2 cm}
\epsfbox{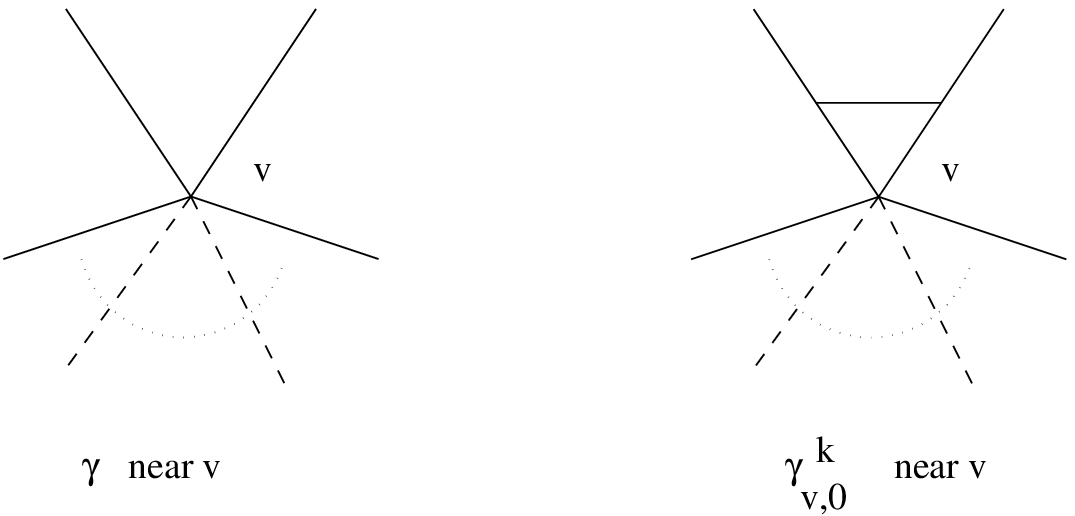}
\vskip .2 cm
{\centerline {Fig. 1}}
\vskip .2 cm
{\noindent} As
a result, $\langle \Gamma, f | \hat{H}(N) = \sum_{\Gamma'} \langle
\Gamma', f^{\Gamma'}|$ where each $\gamma'$ differs from
$\gamma$ by the {\it  removal} of such an edge and for 
$\langle \Gamma, f| \hat H^\dagger(N)
= \sum_{\Gamma'} \langle \Gamma', f^{\Gamma'} |$
each $\gamma'$ differs from $\gamma$ only by the {\it addition}
of such an edge.

There are in fact several types of terms to consider in calculating
the commutator and it is useful to dispense with some of them right
at the beginning.  Suppose that we define the `symmetrized' Hamiltonian
$\hat H^S(N) = \hat H(N) + \hat H^\dagger (N)$.  The commutator
of two such operators can be expanded in four terms:
\begin{eqnarray}
[\hat H^S(N), \hat H^S(M) ] &=& [ \hat H(N), \hat H(M)]
\cr &+& [\hat H^\dagger(N), \hat{H}(M)] + [\hat H(N), H^\dagger(M)]
+ [\hat H^\dagger(N), \hat H^\dagger(M)].
\end{eqnarray}
Since we have already shown that the first term is zero, we shall
concentrate on the last three.  In fact, it is the treatment of 
the middle two that will be most complicated. It is quite
easy to outline the proof that these terms also vanish. Given 
a state $\langle f,\Gamma|$ and a probe spin-network function
$|\Gamma' \rangle $ the quantity $\langle f,\Gamma| 
[\hat H^S(N), \hat H^S(M) ]| \Gamma' \rangle $ is the sum
of contributions coming, roughly speaking, from the action 
of the two operators either at a same vertex $v$, say, of $\Gamma'$
or at disjoint vertices $v$, $w$, say. For the first kind of
contribution, the only 
dependence on the lapse functions is an overall factor 
$N(v)M(v)$ which is obviously symmetric with respect
to the change $(N,M)\mapsto(M,N)$.  For the second,
the action at each of the vertices depends only on the
characteristics of the spin-network $\Gamma'$ in an appropriate 
neighborhood
of that vertex. In most cases the neighborhoods are disjoint 
so that the order in which the operators act is irrelevant.
The only  special case  is when an edge added by the 
operator at $v_1$ can be annihilated at $v_2$ \cite{Simon}.    
On such spin-networks,  the commutator is not zero. 
Thus, in computing the commutator of the constraints, our general approach
will be quite similar to (but dual to) that of section \ref{comsec}.  
That is, we will attempt to decompose the action of $\hat H(N)$ and 
$\hat H^\dagger(N)$ on $\langle \Gamma, f|$ as a sum over actions 
that we may describe as `localized at the various
vertices of $\Gamma$' and then use local diffeomorphism invariance (and
properties of Thiemann's proposal) to compute the commutator.
However, before attacking the problem directly, it will
be useful to recall some general facts about spin networks and cylindrical
functions on $\overline{A/G}$.

We recall that the space $\overline{A/G}$ can in fact be constructed
as a quotient space. In the notation of \cite{MM} 
(see also a later work \cite{ALdiff}), $\overline{A/G} = 
\overline{A}/\overline {G}$.  We will not dwell on the details here, but
simply mention that $\overline{A}$ is a space of distributional
connections much like $\overline{A/G}$, except that holonomies
themselves (and not just their traces) are defined along {\it all}
piecewise smooth curves, open as well as closed.  A certain
gauge group $\overline{G}$ acts on the space $\overline{A}$, and
$\overline{A/G}$ is the quotient $\overline{A}/\overline{G}$.
Thus, the space of functions
on $\overline{A/G}$ is just the space of functions on $\overline{A}$
which are invariant under the gauge group $\overline{G}$.  In particular, 
the gauge invariant spin network functions can be constructed as sums
and products of simpler functions on $\overline{A}$, each of which is
not separately gauge invariant. For example, one may think of the objects
$U^i[\alpha]$ and $|\gamma, j, h_i(\gamma, j, v, I, J)c \rangle$ 
of section \ref{RST} for fixed $i$ as being functions on 
$\overline{A}$\footnote{There is a natural generalization 
of the spin-networks to the `extended spin-networks' defined
in \cite{ALarea}.}.
It will be useful below to `take apart' a spin network function into
a product of several functions that are not separately gauge 
invariant. To this end, suppose that we have a spin network function 
$\Gamma = (\gamma, j,c)$ and an open set ${\cal U} \subset \Sigma$
such that the boundary of ${\cal U}$ does not contain any
vertices of $\Gamma$.
Then there is a (gauge dependent) spin network
function $\Gamma_U$ defined by the triple $(\gamma_U, j_U, c_U)$ such
that the graph $\gamma_U$ is  $\overline U \cap R(\gamma)$ where
$\overline U$ is the closure of $U$. If an edge
$e_U$ of $\gamma_U$ is part of an edge $e$ of $\gamma$, then
$j_U(e_U) = j(e)$ while if $v$ is a vertex of $\gamma_U$, then
$c_U(v) = c(v)$. Furthermore, such edges as oriented in a manner
consistent with the orientation of edges in $\gamma$.
Here, we will call a point $v$ a vertex
of $\gamma_u$ only if it was a vertex of the original graph $v$.  Thus, 
this process does not create new vertices on the boundary of $U$ and
no contractors need be assigned to points on this boundary.  Note
also that the graph underlying a gauge dependent spin network is
open.  The spin network $\Gamma_U$ will also be denoted 
$R(\Gamma) \cap \overline U$.

Strictly speaking, $\Gamma_U$ is not just a function on $\overline{A}$,
but a set of functions labeled by one gauge index $i$ in the spin $j$
representation (or its complex conjugate) 
for every initial (final) end of a spin $j$ edge of $\Gamma_U$ which
is not at a vertex of $\Gamma_U$.  We will call such an end a
`virtual vertex' of $\Gamma_U$.  Given two gauge dependent
spin networks $\Gamma_1$ and $\Gamma_2$, we will say that $\Gamma_1$
is `consistent' with $\Gamma_2$ if, for every initial virtual vertex of
$\Gamma_1$ of spin $j$, it is either a {\it final} virtual vertex of 
$\Gamma_2$ of spin $j$, or it is not a point on the graph $\gamma_2$.
Similarly, final virtual vertices of $\Gamma_1$ which lie in
$\gamma_2$ should be initial virtual vertices, and the virtual
vertices of $\Gamma_2$ should satisfy a similar condition with 
respect to $\gamma_1$ so that this condition is symmetric.
When $\Gamma_1$ is consistent with $\Gamma_2$, there is a 
naturally defined product spin network $\Gamma_1 \Gamma_2 = \Gamma_2
\Gamma_1$ given by multiplying the associated functions on 
$\overline{A}$ and contracting any indices that correspond to the
same virtual vertex. Note that when $U$ is related to a
(gauge invariant) spin network $\Gamma$ as above, we have
$\Gamma = \Gamma_U \Gamma_{(\overline U)^c}$, where
${}^c$ denotes the complement of a set. 

We now  commence the proof by finding a more convenient
expression for $ \langle \Gamma, f | H^\dagger(N)$.  
In particular, note
that if $\langle \Gamma', f |H (\tilde N) |\Gamma \rangle$
is nonzero, then $|\Gamma' \rangle$ is diffeomorphic to a spin network
over a graph $\gamma_0$ of which $\gamma$ is a subgraph $(\gamma_0 \supset
\gamma)$.  In fact, $\gamma_0$
differs from $\gamma$ only
by having a single extra edge.   We would like to
fix the position of this extra edge in some sense,
but to keep it close to $v$.

As a result, we introduce a special set $S^{\Gamma_U, \dagger}_{v,U}$.
Consider a spin network $\Gamma$ and an open set $U$ as above, such
that $U$ contains exactly one vertex $v \in V(\Gamma)$.
Furthermore, the set $\overline{U} \cap R(\gamma)$ should
be connected, where again
$\overline{U}$ denotes the closure of $U$.  Now, choose
an arbitrary function $\tilde N$ whose support includes $v$ but includes
no other vertex of $\Gamma$.

As indicated, the set $S^{\Gamma_U,\dagger}_{v,U}$ will
depend only on the vertex $v$, the 
set $U$, and the spin network $\Gamma_U$ defined by the part of $\Gamma$
inside $\overline U$.  This is to be some fixed set
of (gauge dependent) spin networks $\Gamma'_U = (\gamma'_U,j'_U,c'_U)$
such that A) each spin
network is entirely contained in $\overline
U$: $\gamma'_U \subset \overline U$, B) for
and $\Gamma_1$ with $R(\Gamma_1) \cap \overline U$ = $\Gamma_U$, 
$\langle \Gamma_0, 1 | H(\tilde N) | \Gamma_1 \rangle$ is nonzero only
if $\Gamma_0$ is diffeomorphic to a state $\Gamma_0'$ such that
$R(\Gamma'_0) \cap \overline U$ is a nontrivial member of the span $\overline 
S^{\Gamma_U,\dagger}_{v,U}$ of $S^{\Gamma_U,\dagger}_{v,U}$;
$\varphi(R(\Gamma_0)) 
\cap \overline U \in
\overline S^{\Gamma_U,\dagger}_{v,U}$ for some $\varphi \in Diff(\Sigma)$.
We may also choose $S^{\Gamma_U,\dagger}_{v,U}$
to be independent of $\tilde N$.  
Note that, the 
spin assignments of $\Gamma_1$ and $\Gamma_0$ can only differ by a bounded
amount. Therefore, since a finite set of graphs suffices, 
we may choose the set $S^{\Gamma_U, \dagger}_{v,U}$ to be finite.

Consider now the action of the 
hermitian conjugate operator $H^\dagger(1)$ on
a state $\langle \Gamma, f |$.  This can be expanded as
\be
\langle  \Gamma, f| H^\dagger(1) = \sum_{v \in V(\Gamma)} \sum_{\Gamma_U'
\in S^{\Gamma_U, \dagger}_{v,U}} \langle \Gamma'_U \Gamma_{( \overline
U)^c}, f \circ i^*_{\Gamma,U, \Gamma_U'} | a^{\Gamma_U,\Gamma_U'} 
\ee
for an appropriate set of coefficients $a^{\Gamma_U,\Gamma_U'}$
and where $i^*_{\Gamma,U, \Gamma_U'} : \Sigma^{V(\Gamma_U'\Gamma_{U^c})}
\rightarrow  \Sigma^{V(\Gamma)}$ is the map on assignments of points
to vertices induced by the embedding of $V(\Gamma)$ in $V(\Gamma_U'
\Gamma_{(\overline U)^c})$.  That is, for $\sigma \in  
\Sigma^{V(\Gamma_U'\Gamma_{(\overline U)^c})}$,  we have
$i^*_{\Gamma, U, \Gamma_U'} (\sigma)
= \sigma|_{V(\Gamma)}$.  Note that $V(\Gamma)$ is always
a subset of $V(\Gamma'_U \Gamma_{(\overline U)^c})$.
To fully define the coefficients
$a^{\Gamma_U, \Gamma_U'}$, suppose that $\Gamma_0$ is consistent
with $\Gamma_U$ and $\Gamma'_U$, satisfies
$\gamma_0 \subset U^c$, that both $\Gamma_U \Gamma_0$ and
$\Gamma'_U \Gamma_0$ are gauge invariant spin networks over graphs
with trivial symmetry groups, and that $\Gamma_U \Gamma_0$ has
only one vertex.  Such a $\Gamma_0$ always exists, and we define
\be
a^{\Gamma_U,\Gamma_U'} := \langle \Gamma_U, \Gamma_0| H^\dagger(1)
| \Gamma'_U \Gamma_0 \rangle.
\ee
The result is independent of the choice of $\Gamma_0$.

To include a nontrivial lapse function, we need just a bit more
notation.  For a function $f : \Sigma^{V(\Gamma)} \rightarrow {\bf C}$, a
vertex $v \in V(\Gamma)$, and
a function $N: \Sigma \rightarrow {\bf C}$, let the product
$f \star_v N : \Sigma^{V(\Gamma)} \rightarrow {\bf C}$ be the function
\be 
(f \star_v N) (\sigma) = f(\sigma) N(\sigma(v)).
\ee
This product is commutative in the sense that, given another function
$M: \Sigma \rightarrow {\bf C}$, we have
\be 
\label{starcom}
f \star_v N \star_{v'} M =
f \star_{v'} M \star_v N.
\ee
The action of a general $\hat H(N)$ on $\langle
\Gamma, f|$ is then
\be
\label{long}
\langle  \Gamma, f| H^\dagger(N) = \sum_{v \in V(\Gamma)} \sum_{\Gamma_{U_v}'
\in S^{\Gamma_{U_v}, \dagger}_{v,{U_v}}} \langle \Gamma'_{U_v} \Gamma_{
(\overline {U_v})^c},
f \circ i^*_{\Gamma,U_v, \Gamma_{U_v}'} \star_v N | a^{\Gamma_{U_v},
\Gamma_{U_v}'}. 
\ee
where appropriate open sets $U_v$ have been chosen.
We now condense this slightly by defining, for each $v \in V(\Gamma)$, 
operators $H^\dagger_v(N)$ given by
\be
\label{local}
\langle  \Gamma, f| H_v^\dagger(N) := \sum_{\Gamma_U'
\in S^{\Gamma_U, \dagger}_{v,U}} \langle \Gamma'_U \Gamma_{\overline(U)^c},
f \circ i^*_{\Gamma,U, \Gamma_U'} \star_v N | a^{\Gamma_U,\Gamma_u'}. 
\ee
Note that definition of $H^\dagger_v(N)$ does not depend on the open set $U$.

Unfortunately, it is not really correct to call $H^\dagger_v(N)$ an
operator on states: if the same state is parameterized in two different
ways ($\langle \Gamma, f| = \langle \Gamma', f'|$), then (\ref{local})
in general gives different results.  The object $H^\dagger_v$ is
more properly considered as an operator on certain lists of pairs
$(\Gamma, f)$.  Nonetheless, it will be convenient to treat $H^\dagger_v$
as a operator and to not introduce further notation to treat it properly.
So long as the true character of this object is understood, this should
not cause any problems.

Another useful property of $\hat H^\dagger_v(N)$ for $v \in V(\Gamma)$
is that $\hat  H^\dagger_v(N)$ annihilates the state $\langle \Gamma, f|$
unless the vertex $v$ in $\Gamma$ has at least three incident edges
with linearly independent tangents;
we shall call such vertices `eternal'
since they are neither added nor removed by the action of the
Hamiltonian constraints (see figure 1).
Furthermore, we may note that,
for $\Gamma'_U \in S^{\Gamma_U,\dagger}_{v,U}$ we have
$V(\Gamma'_U) \supset V(\Gamma_U)$ and that  any vertex
present in any $\Gamma_U'$ which is not in $\Gamma_U$ fails to
have three incident edges with independent tangents.  Thus, if we
act again with another $\hat H^\dagger(M)$ and expand the result as above, 
only terms involving $\hat H^\dagger_{v'}(M)$ where $v'$ is
a vertex of the original graph $\Gamma$ are nonzero.

This notation can be now 
used to calculate the commutator $[ \hat H^\dagger(N), 
\hat H^\dagger(M)]$ very directly.   As described above, 
\be
\label{vertexp}
\langle \Gamma, f| [\hat H^\dagger(N), \hat H^\dagger(M)] = \sum_{v,v'
\in V(\Gamma)} \langle \Gamma, f| (H_v^\dagger(N) H_{v'}^\dagger(M)
-   H_{v'}^\dagger(M) H_v^\dagger(N)  ).
\ee
Now, let $U \ni v$ and $U \ni v'$ be disjoint open sets such that the closures
$\overline {U}$, $\overline{U'}$ each contain
exactly one vertex ($v$ or $v'$) of $\Gamma$ and such that each 
graph $R(\gamma) \cap \overline U$
and $R(\gamma) \cap \overline {U'}$ is connected.  
Then, for each spin network $\Gamma'_{U} \Gamma_{(\overline
U)^c}$  in (\ref{local}), 
$R(\Gamma'_{U}) R(\Gamma_{(\overline U)^c}) \cap \overline{U'}
= R(\Gamma) \cap \overline{U'} =: R(\Gamma_{U'})$ is independent of 
$\Gamma'_U$.
Thus, the operator
$H^\dagger(M)$ acts on each such term in essentially the same way and
we have:
\be
\label{twodag}
\langle
\Gamma, f | H^\dagger_v(N) H^\dagger_{v'}(M) = \sum_{\Gamma'_U 
\in S^{\Gamma_U, \dagger}_{v, U}} \sum_{\Gamma'_{U'} \in
S^{\Gamma_{U'}, \dagger}_{v',U'} } \langle \Gamma'_U \Gamma'_{U'}
\Gamma_{(\overline{U \cup U'})^c} , f \circ i^* \star_v
N \star_{v'} M| a^{\Gamma_U, 
\Gamma'_U} a^{\Gamma_{U'}, \Gamma'_{U'}}.
\ee
Here, we have used $i^*$ to denote each of the maps 
$i^*_{\Gamma, U, \Gamma'_U} \circ i^*_{\Gamma'_U \Gamma_{(\overline U)^c},
U', \Gamma'_{U'}}$ which should appear in (\ref{twodag}).  The point
is that all of these maps act in essentially the same way: they
simply restrict an assignment $\sigma : V(\Gamma'_U \Gamma'_{U'} 
\Gamma_{(U \cup U')^c}) \rightarrow \Sigma$ to $V(\Gamma)$.
In writing (\ref{twodag}), we have used the fact that composition
with $i^*$ and the star product over $v$ commute when $v \in V(\Gamma)$;
that is,
\be
\label{vi}
(f
\star_v N) \circ i^* = (f \circ i^*) \star_v N
\ee
since $[i^*(\sigma)](v) = \sigma(v)$.
Due to the commutativity (\ref{starcom})  of the star product, 
it is clear that $\langle \Gamma, f| \hat H^\dagger_v(N) \hat H^\dagger_{v'}
(M) = \langle \Gamma, f| \hat H^\dagger_{v'}(M) \hat H^\dagger_{v}(N)$.
Using (\ref{vertexp}) above, we find that $[\hat H^\dagger(N),\hat
H^\dagger(M)]=0$.

We would now like to define  `localized' versions of the
operators $\hat H(N)$ in analogy with the definition of $\hat H^\dagger_v(N)$.
This will allow us to compute the terms $[\hat H^\dagger(N),
\hat H(M)]$ and $[\hat H(N), \hat H^\dagger(M)]$.
This time, however, there will be certain differences.  Recall
that the action of $\hat H(N)$ on a state $\langle \Gamma, f |$
generates a series of terms $\langle \Gamma', f'|$ such that
each $\Gamma'$ sits over some graph $\gamma'$ given by {\it removing} some
edge from $\gamma$.  As a result, it will be useful to decompose
the action of the constraint operator not only with respect to
the `vertex at which it acts,' but also with respect to the
`edge that is removed'.  We denote the set of edges of $\Gamma$ by
$E(\gamma)$.

Let us begin with the following observation. 
Consider any edge $e$ of $\Gamma$ and let $U$ be any open set
which contains $e$ and $v$, as well as any edges which connect
$v$ to the endpoints of $e$.  However, the closure $\overline{U}$
of $U$ is not to contain any other vertices and $\gamma \cap
\overline{U}$ is to be connected.
Now, consider the set $\tilde S^{e,\Gamma_U}_{v,U}$ of (gauge dependent)
spin networks such that
for $\Gamma'_{U} \in S^{e,\Gamma_U}_{v,U}$ we have A) the spin network is
fully contained in $\overline U$:
$\Gamma'_U \cap \overline U = \Gamma'_U$, B) the 
associated graph
$\gamma'_U$ is the subgraph of $\gamma_U$ obtained by removing
the edge $e$ and C) $\langle \Gamma, 1 |
\hat H(\tilde N) |\Gamma'_U \Gamma'_{(\overline U)^c}
\rangle$ is nonzero for some
$\Gamma'_{(\overline U)^c}$ consistent with $\Gamma'_U$
and some function $\tilde N$ whose support 
${\rm supp} \tilde N$ satisfies ${\rm supp } \tilde N \subset U$ and
${\rm supp} \tilde N \cap V(\Gamma_U) = \{ v\}$.
Let $\overline{S}^{e,\Gamma_U}_{v,U}$ be the linear space spanned
by $\tilde S^{e, \Gamma_U}_{v,U}$ and let $S^{e,\Gamma_U}_{v,U}$
be a basis for that space.  Because the change in spins
caused by $\hat H (\tilde N)$ is bounded, and because there are a finite
number of subgraphs of $\gamma$, the set $S^{e,\Gamma_U}_{v,U}$
is finite.  Furthermore, if
$\langle \Gamma, 1 | \hat H(\tilde N)
|\Gamma_0 \rangle$
is nonzero, then  for some $\varphi \in Diff(\Sigma)$, 
$\varphi (\Gamma) \cap \overline U$ 
is a nontrivial element of the space 
$\overline S^{e, \Gamma}_{v,U}$ for some edge $e$. 

For each vertex $v$ and spin networks $\Gamma', \Gamma$ for which
$V(\Gamma) \supset V(\Gamma') \ni v$, let us introduce the maps
$\eta^{*, \Gamma, \Gamma'}_{v}: \Sigma^{V(\Gamma')} 
\rightarrow \Sigma^{V(\Gamma)}$ and
$\eta_v^{\Gamma', \Gamma}: V(\Gamma) \rightarrow V(\Gamma')$
defined by
\begin{eqnarray}
\eta^{\Gamma', \Gamma}_v (v') &=& \Bigg\{ {
{v \ {\rm if } 
\ v' \notin V(\Gamma') } 
\atop
{v' \ {\rm if } \ v' \in V(\Gamma')}
} \ \  \ {\rm and} \cr
\eta^{*,\Gamma', \Gamma}_v(\sigma) &=& \sigma \circ \eta_v^{\Gamma',\Gamma}.
\end{eqnarray}
The action of $\hat H(N)$ on $\langle \Gamma, f|$ may then be written
\be
\langle  \Gamma, f | \hat 
H(N) = \sum_{e \in E(\gamma)} 
\sum_{v \in V(\Gamma)} \ \ \sum_{\Gamma'_{U_{e,v}}
\in S^{e,\Gamma_{U_{e,v}}}_{v,U_{e,v}}}
\langle  \Gamma'_{U_{e,v}} \Gamma_{U_{e,v}^c}, f \circ \eta^*_v \star_v N | 
b^{\Gamma'_{U_{e,v}}, \Gamma_{U_{e,v}}}
\ee
where $E(\gamma)$ are the edges of $\gamma$ and an appropriate
collection of open sets $U_{e,v}$ has been chosen and we have 
used $\eta^*_v$ as an abbreviation for $\eta_v^{*, \Gamma'_{U_{e,v}}
\Gamma_{U^c_{e,v}}, \Gamma}$.  The coefficients 
$b^{\Gamma'_{U_{e,v}}, \Gamma_{U_{e,v}}}$ are given by
\be
\label{bs}
b^{\Gamma'_{U_{e,v}}, \Gamma_{U_{e,v}}} =
\langle \Gamma_0 \Gamma_{U_{e,v}} | H(1) | \Gamma_0 \Gamma'_{U_{e,v}}
\rangle
\ee 
for some gauge dependent spin network $\Gamma_0$ contained in
$(\overline {U_{e,v}})^c$ consistent with $\Gamma_{U_{e,v}}$ and 
$\Gamma'_{U_{e,v}}$ such that $\Gamma_0 \Gamma_{U_{e,v}}$ and $\Gamma_0
\Gamma'_{U_{e,v}}$ are gauge invariant spin networks over graphs with
no symmetries, and such that the only eternal
vertex of $\Gamma_0 \Gamma_{U_{e,v}}$ 
is $v$.  Such a $\Gamma_0$ always exists, and  (\ref{bs}) does not
depend on the particular choice of $\Gamma_0$.

In analogy with $H^\dagger_v(M)$, we now introduce an object $H_{e,v}(N)$.
Given $\langle \Gamma, f|$ with $e \in E(\Gamma)$ and $v\in
V(\Gamma)$ and given
an open set $U$ as in the definition of $S^{e,\Gamma_U}_{v,U}$
above, we define
\be
\langle \Gamma, f |\hat H_{e,v}(N)
:= \sum_{\Gamma'_U \in S^{e,\Gamma_U}_{v,U}} \langle
\Gamma'_U \Gamma_{(\overline U)^c},
f \circ \eta^*_v \star_v N | b^{\Gamma'_U,
\Gamma_U}
\ee
so that 
\be
\label{Hact}
\langle \Gamma, f | \hat H(N) = \sum_{e \in E(\Gamma)} \sum_{v \in
V(\Gamma)} \langle \Gamma, f | \hat H_{e,v} (N).
\ee
as before, the state (\ref{Hact}) does not depend on the choice of the
open set $U$.

Suppose that we wish to act on (\ref{Hact}) with $H^\dagger_{v'}$
for some $v' \in V(\Gamma)$, $ v \neq v'$.  Recall that we need only
act at eternal vertices $v'$ which were present in the original 
graph $\gamma$.
Since $v \neq v'$, this means that 
$v' \in V(\Gamma_{(\overline U)^c})$
and that the action of $\hat H^\dagger_{v'}(M)$
on each term is well-defined.  Choosing $U' \cap U = \emptyset$ and
$U'$ as in the definition of $S^{\Gamma_{U'} , \dagger}_{v,U'}$, it
is clear that the action of $\hat H^\dagger_v(M)$ is decoupled from that
of  $\hat H_{e,v}(N)$ and we have
\be
\label{01}
\langle \Gamma, f | H_{e,v}(N) H^\dagger_{v'}(M)
= \sum_{\Gamma'_U \in S^{e,\Gamma_U}_{v,U}} \sum_{\Gamma'_U \in
S^{\Gamma_{U'},\dagger}_{v,U'}} \langle \Gamma'_U \Gamma'_{U'}
\Gamma_{(\overline{U \cup U'}
)^c}, (f \circ \eta^*_v \star_v N) \circ i^*
\star_{v'} M |.
\ee

On the other hand, for $e\in E(\Gamma)$, $v \in V(\Gamma)$,  we
may act on (\ref{local}) with $\hat{H}_{e,v}(N)$ and the result is
\be
\label{10}
\langle \Gamma, f | H^\dagger_{v'}(M) H_{e,v}(N)
= 
\sum_{\Gamma'_{U'} \in S^{\Gamma_{U'}, \dagger}_{v,U'}}
\sum_{\Gamma'_U \in S^{e, \Gamma_U}_{v,U}}
\langle \Gamma'_U \Gamma'_{U'} \Gamma_{(\overline{U \cup U'})^c}, 
(f \circ i^* \star_{v'} M) \circ \eta^*_v \star_v N |
\ee
where we again used the available freedom to choose $U'
\cap U = \emptyset$.  

Recall (\ref{vi}) that composition with $i^*$ commutes with the star
product over $v' \in V(\Gamma)$.  Also,
so long as $v, v'$ are `eternal' vertices 
that are neither added nor removed by the Hamiltonian constraints
$ f \circ \eta^*_v 
\star_{v'} M = (f \star_{v'} M) \circ \eta^*_v$ (whether 
$v=v'$ or not) since $[\eta_v^*(\sigma)](v')
= \sigma (v')$.  Finally, we note that 
since
\be
\eta^{(\Gamma'_{U'} \Gamma'_U \Gamma_{(U \cup U')^c}), (\Gamma'_{U'}
\Gamma_{(U')^c})}_v |_{V(\Gamma)} 
= \eta^{\Gamma'_U\Gamma_{U^c}, \Gamma}_v,
\ee
we may drop the superscripts on $\eta_v$ and write
\begin{eqnarray}
(i^* \circ \eta^*_v )(\sigma) &=& \sigma \circ \eta_v \cr
&=& \sigma|_{V(\Gamma)} \circ \eta_v = (\eta^*_v \circ i^*) (\sigma) .
\end{eqnarray}
Combining this with commutativity (\ref{starcom}) of the star product, 
we see that (\ref{01}) and (\ref{10}) are identical for eternal
vertices $v$ and $v'$.

However, this 
does {\it not} imply that $[\hat H(N), \hat H^\dagger(M)] =0$.
The point is that, while summing (\ref{01}) over
$v,v' \in V(\Gamma)$ and $e \in E(\Gamma)$ gives
$\langle \Gamma, f |
\hat H(N) \hat H^\dagger(M)$, it is not true that summing (\ref{10})
over this set gives
$\langle \Gamma, f | \hat H^\dagger(M) \hat H(N)$.  This is because, 
when $\hat{H}(N)$ acts on (\ref{local}), it generates terms corresponding
not only to edges of the original graph $\Gamma$, but also to the edges
(that we will call $e(\Gamma'_U,\Gamma_U)$) which were created by
$\hat H^\dagger(M)$.  Thus, there are still terms of the form
\be
\label{badterm}
\sum_{\Gamma'_{\overline{U'}} \in S^{\Gamma_U, \dagger}_{U,v}}
\langle \Gamma'_{U'} \Gamma_{(\overline{U'})^c},
f\circ i^* \star_{v'} M \star_{v} N
| a^{\Gamma_{U'}, \Gamma'_{U'}} \hat{H}_{v',e(\Gamma_U,
\Gamma'_U)}
\ee
to consider.  For the special case $v= v'$, terms of this type can give
no net contribution as it is clear that each such term is symmetric
in $M$ and $N$, whereas the commutator $[\hat{H}^S(N), \hat{H}^S(M)]$
is antisymmetric.  The important question is therefore whether there
are any terms of this kind for $v \neq v'$.  

Let us therefore examine the requirements for the existence of such a term.
The operator $\hat H^\dagger(M)$ creates an edge which is such that it
can be `slid' to the vertex $v$; that is, the new vertices
created are on edges $e_1,e_2$
which are incident at $v$, and there are no
vertices along $e_1,e_2$ between $v$ and the new vertices (see
figure 1).  In the terms of interest, this newly created edge
is then removed by the action of $ \hat H_{v'}(N)$ acting at some
{\it other} vertex $v'$.  However, $\hat H_{v'}(N)$ can only remove
edges that may be `slid' to $v'$.  Thus, such a term can only
arise when the graph $\gamma$ has a subgraph (Simon's subgraph \cite{Simon}) 
of the form
\vskip .2cm
\centerline {\epsfbox{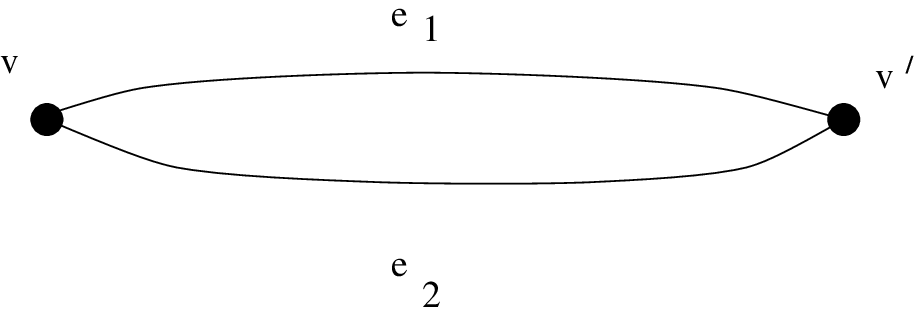}}
\vskip .2cm
{\centerline {Fig. 2}}
\vskip .2cm
{\noindent}
with no additional edges linking with the above subgraph.

Explicit computations which we have made using the detailed form of
the coefficients of Thiemann's Hamiltonian \cite{DPR}
show that, when $\gamma$ is of this type, 
$\langle \Gamma, f | [ H^S(N), H^S(M)]$ does not in general vanish.
However, this has nothing to do with whether or not the function $f$ is
diffeomorphism invariant: 
even for constant functions such as $f=1$, 
the action of the commutator on a diffeomorphism invariant state
 $\langle \Gamma, 1|$ is nonzero when $\Gamma$ is based on a graph
$\gamma$ containing Simon's subgraph.  Thus, the commutator of the 
symmetrized Hamiltonians may be called anomalous on such states.

In summary then, the space $\T'_*$ splits into two subspaces
$\T'_* = \T'_0 \oplus \T'_A$ where $\T'_0$ is built from states $\langle
\Gamma, f|$ such that $\gamma$ has no subgraph of the form described by
figure 2, while $\T'_A$ is built from those that do.  It is worth
mentioning that there are many subspaces of $\T'_0$ that are invariant
under the action of $H^S$, so that it would be possible to simply
restrict the definition of $H^S$ to such a subspace.  The commutator
$[\hat H^S(N),\hat H^S(M)]$ annihilates all states in $\T'_0$, but does
not annihilate the general state in $\T'_A$.
For the case of the unprojected constraints on $\T'_{*,2}$, the
result is similar except that the corresponding $\T'_A$ is larger
and in the associated $\T'_0$ the only subspaces invariant under
the action of $\hat H(N)$ and $\hat H^\dagger(N)$ are rather
small.

Finally, we note that the need to split $\T'_*$ into the spaces
$\T'_0$ and $\T'_A$ is a consequence of the particular loop assignment
chosen above.  If, on the other hand, only loops $\alpha(\gamma, v,I,J)$
that did not overlap
the original graph  $\gamma$ (and therefore the intersect $\gamma$
only at the vertex $v$) were used instead, an analogous argument
could be made but, this time, there would be no terms of the form
(\ref{badterm}).  Instead, we would have $[\hat H^S(N), \hat H^S(M)] = 0$
on all of $\T'_*$.

\section{Discussion}

In this work, we considered `RST-like' proposals for the 
the Euclidean 
Hamiltonian constraints of quantum gravity in a loop representation
which follow the suggestions of \cite{CL} and define the operators
through a particular kind of limiting construction.  We have shown that
this limit converges
not only on the diffeomorphism invariant states $\T'_{Diff}$
where it was first introduced, but in fact on a much larger space of
`vertex-smooth' states,
$\T'_* \supset \T'_{Diff}$.  In particular, $\T'_*$ contains the space in
which the `physical states' were sought in \cite{I}.  One might also
have liked to introduce an inner product on $\T'_*$.  We have not done so, 
and in fact the appendix shows that a fully satisfactory inner product 
cannot be introduced on {\it any} space
in which Hamiltonian-like constraints are well defined and which
contains a diffeomorphism 
invariant state not annihilated by the constraints.
The results of the appendix may  have consequences for the more
general idea 
of first solving the diffeomorphism constraint and then
defining the Euclidean Hamiltonian constraint on the resulting solutions.

We have also computed the algebra of RST-like constraints.  For
any two members of a family of such operators, their commutator
is identically zero on $\T'_*$.  Furthermore, we have addressed 
suggestions for `symmetrizing' constraints of the type described in
\cite{I,II,DPR},
with the result that their commutator 
is again identically zero.  However, because the proposed
`symmetrizations' are considered unsatisfactory for other 
reasons, we also introduced another symmetrization which
seems to be of the type desired by many researchers.  With
this latter definition, we find that $\T'_*$ decomposes
as $\T'_0 \oplus \T'_A$ with the commutator vanishing on $\T'_0$
and being anomalous (in the sense defined above) on $\T'_A$.  Thus, we
have addressed all existing proposals for the Hamiltonian constraints
in loop quantum gravity which involve a limiting procedure in which
loops are shrunk to a point.  Such results are in agreement with
the prediction of \cite{BB} that, if a procedure for removing the
regulators in a loop representation would be found, the resulting operators
would not satisfy the classical algebra.  The only proposal known to
us which is not of the type considered here is that of \cite{Lee},
and this will
be addressed in \cite{GLMP}.

Although we have discussed only the Euclidean constraints here, 
an extension of the results of section II to the Lorentzian
case is straightforward.  The only definition of the Lorentzian constraints
is that given in \cite{I}, and
the proposed Lorentzian constraints are constructed in much the
same way as the RST-like operators.  By rewriting the `anomaly-free'
calculation of \cite{I} along the lines of our section \ref{comsec}, 
it is clear  that the commutator
of two such Lorentzian Hamiltonian constraints vanishes as well.

This calculation was intended as a test of the extent to which 
the proposed quantum constraints capture the classical structure
of general relativity. We found that their algebra is correct
as long as the commutators are applied to diffeomorphism invariant
states. At the level of diffeomorphism non-invariant states,
the answer was found to be `rather little.'  It would
appear that three interpretations are possible:

  The first would be to
note that the constraint algebra is not actually a `physical' object since
the constraints should simply annihilate any physical states.  Thus, 
a consistent interpretation is to state that
point of view to state that any meaningful comparison of a
quantum theory with general relativity must take the
form of examining the classical limit of
gauge invariant quantities and physical states.  Since our analysis
does not achieve this, it is not truly meaningful.

As a first response, let us consider a theory of gravity coupled to matter.
Then the classical phase space functions corresponding to the operators
studied here no longer annihilate physical states and neither does
their commutator.  Instead, while that commutator is not an observable, 
it does generate `diffeomorphisms
of the gravitational degrees of freedom relative to the matter degrees
of freedom' even on physical states.
However, if our results carry over to such a setting and if the physical
states lie in a space corresponding to the one studied here, then
the analogous quantum commutator is still the zero operator.   
It is true that even this would not address the action of
a physical operator on a physical state.  Nonetheless,
it appears close enough that
our work may be taken as a caution that, when confronted with a
a proposed quantum object arising from a complicated
regularization procedure, it is important to find a nontrivial
check that this object does in fact have some relation to the desired
physics. 

The next interpretation would be to suggest that the space
$\T'_*$ is somehow too small to see a nontrivial commutator;
that the commutator just `happens to vanish' on $\T'_*$.  In argument
against this interpretation we note that
$\T'_*$ is quite large and contains both the spaces $\T$ and $\T'_{Diff}$, 
which were supposed to capture much of the physics of general 
relativity\cite{weaves,RSarea,ALarea,ALvolume}.
However, 
this question merits a more thorough
discussion of the `right-hand side' of the classical
commutator.
Recall that the classical Poisson bracket of two Hamiltonian constraints
is
\be
\{H(N),H(M)\} = \int N_a C_b q^{ab}, 
\ee
where $q^{ab}$ is the inverse three-metric and 
$C_b$ is the vector constraint.  The question is then
of how  $\int N_a C_b q^{ab}$ should be represented on $\T$ or $\T'_*$ and
whether or not it should, in general, vanish.\footnote{This
issue was first raised with us by Jorge Pullin.}.
One study of this operator
was performed in \cite{III} and it will be addressed further in 
\cite{GLMP}, but here we content ourselves with two observations.
We recall that $C_b$
generates diffeomorphisms and that the generator of diffeomorphisms is
well-defined and nonvanishing on $\T'_*$.  Also, $q^{ab}$ is invertible 
classically, so we would be surprised if it vanished on a large set of 
quantum states.  

The third interpretation is that the quantum constraint proposals studied
here fail to capture the physics of general relativity.  
Taking that point of view the question remains,  however, 
of whether one can single out a particular aspect of these proposals
as being responsible for the difficulties.  The use of diffeomorphism 
invariant
(or partially diffeomorphism invariant) states in defining the limit in which
the regulators are removed seems a likely culprit.  This question will
be examined in more detail in \cite{GLMP}, which will study certain variations
on that theme.

It is, however, important to note that even if these constraints
by themselves fail to capture the physics of gravity, this does not
necessarily mean that they are {\it incompatible} with that physics
and cannot be used as a starting point.  An important example in this regard
is the work of Thiemann \cite{T2+1} on 2+1 gravity.  We expect that
all of the results given here in the 3+1 context also hold for the
2+1 constraints used there.  Nevertheless, Thiemann showed that the solutions
space of these constraints contains (as a small subspace) the usual
physical states of the Witten formulation \cite{Witten} of 2+1 gravity.
He was then able to pick out these states by using various elements
of the Witten formulation.  Something similar may be possible in the
3+1 case as well, though the question remains of what additional
structure would then play the role of the 2+1 Witten formulation.

It would appear that our work is related to the commentary of
Smolin in \cite{Lee}.  His work notes that constraints
defined by the RST-like limiting procedure (and their Hermitian conjugates)
are `too local' at an
intuitive level and do not seem to generate structures that resemble
the features of general relativity.  It then suggests a number of
difficulties which may be expected to follow, but unfortunately
it is difficult to make these arguments conclusive.
While the algebra was not a specific concern of \cite{Lee}, 
at least for the RST-like operators
themselves (if not for the `symmetrized' versions of section \ref{sym}),
it is the intuitive `locality' which is responsible for the vanishing of 
the commutator.  
Thus, although our analysis does not deal directly with observables, 
one could regard our work in section \ref{RST}
as a precise statement of the type desired in \cite{Lee}.

One might hope \cite{RR,LF}
to achieve better results by abandoning the canonical
framework completely and defining the quantum theory via some sort of
path integral.  While this may be possible, we take the results
derived here as a general warning that, until some well-defined and nontrivial
calculation can be done to connect the quantum theory with 
general relativity, it remains unclear to what extent the
proposed quantization captures the desired physics.

\acknowledgements

We would like to thank Abhay Ashtekar, Rudolfo Gambini, Jos\'e Mour\~ao, 
Jorge Pullin, Carlo Rovelli, and especially Thomas Thiemann both for their
perspective and for clarifying the status of the field.  We also
thank Thomas Thiemann for a critical reading of an early draft of the paper.
This was work supported in part by funds from the
Erwin Schr\"odinger Institute, the Max Planck Society, the grants
PHY95-07065 and PHY-9722362  from the US National Science foundation, 
Syracuse
University. JL thanks Alexander von Humboldt-Stiftung (AvH),
the Polish Committee on Scientific Research (KBN, grant no. 
2 P03B 017 12)  and the Foundation for Polish-German
cooperation with funds  provided by the Federal Republic of Germany
for the support.
  We would also like to thank both the Erwin Schr\"odinger Institute 
and the Albert Einstein
Institute (Max Plank Instit\"ut f\"ur Gravitationsphysik) for their
hospitality, as much of the work was done while J.L. and D.M. were visiting
these facilities.

\appendix

\section{On an inner product}

In this appendix, we derive the following theorem:
\medskip

\noindent {\bf Theorem:} {\it Suppose that there is a Hilbert space 
${\cal H}$ on which the diffeomorphism
group $Diff(\Sigma)$ acts unitarily through operators ${\cal D}_\varphi$ for
$\varphi \in Diff(\Sigma)$. Here, $\Sigma$ is some compact manifold.
Furthermore, suppose that there
is a family of operators $H(N)$ labeled by smooth real-valued functions on 
$\Sigma$
such that $H(N)$ is linear in $N$, and such that $D_\varphi H(N) D_\varphi^{-1}
= H(N \circ \varphi)$.  Then a
state $|\psi_{Diff} \rangle$ can belong to the common domain of all operators
$H(N)$ only if it is annihilated by them.}  
\medskip

Note that we have not required $H(N)$ to be symmetric or self-adjoint
and that by definition  laps functions have density weight zero. 
For a non-compact manifold $\Sigma$ the same conclusion holds 
provided $H(N)$ is `super linear' with respect to a sum given
by a partition of the unity. That assumption is satisfied by
the RST-Hamiltonian operators. 

Since we have a large class of operators (the RST-like constraints) which
are well-defined on $\T'_{Diff} \subset \T'_*$ and not all of $\T'_{Diff}$ is
annihilated by the constraints, this theorem will show
that there can be no Hermitian
inner product on $\T'_*$ such that the action of $Diff(\Sigma)$ is unitary, 
unless it is very degenerate.

To prove the theorem, consider the inner product  $(H(N)|\psi_{Diff}\rangle,
H(M)|\psi_{Diff} \rangle) \equiv (N,M)$ and note that $(N,M)$ is
a positive definite {\it real} bilinear product on smooth real
functions $N$ and
$M$.  This product is diffeomorphism invariant in the sense that
$(N,M) = (N \circ \varphi,M \circ \varphi)$ for any diffeomorphism
$\varphi \in  Diff(\Sigma)$.  

Let us begin by covering $\Sigma$ with coordinate charts ${\cal U_i}$.  
It is enough to show that $H(N)|\psi_{Diff}\rangle = 0$ for any $N$ 
supported in some fixed coordinate chart ${\cal U}$, since, due to  
the compactness of $\Sigma$,
the general case then follows from diffeomorphism invariance and linearity.

Let $N$ be supported in a chart $\cal U$. Then, there exists
a function $M$ supported in ${\cal U}$ such that, for sufficiently
large value of a parameter $\lambda$, the function $M+{N\over \lambda}$ 
is diffeomorphism
equivalent to $M$. Indeed let $M$ be any smooth function which coincides with 
$x_1$ on $U$.  Since $N$ is smooth and $\Sigma$ is compact, the first
derivative ${{\partial N} \over {\partial x_1}}$
must be bounded by some positive real number $\lambda'_0$.
Thus, for $\lambda > \lambda'_0$, $x_1 + N/\lambda$ is a smooth monotonically
increasing function of $x_1$.
Since $N$ must vanish at the boundary of ${\cal U}$,  $x + N/\lambda = x_1$
on the boundary of ${\cal U}$ and the map $\varphi: (x_1,x_2,...,x_k) \mapsto 
(x_1 + N/\lambda,x_2,...,x_k)$ is a diffeomorphism.  From the diffeomorphism
invariance of the product $(\ ,\ )$, for $\lambda > \lambda_0$,
we must have $(M,M) = (M_\lambda,M_\lambda) = (M,M) + \lambda^{-1} [(M,N)
+ (N,M)] + \lambda^{-2} (N,N)$.  Thus, $(N,M) + (M,N)$ and $(N,N)$ must
vanish.  Since $(N,N)$ is just the norm of $H(N)|\psi_{Diff}\rangle$
in ${\cal H}$, we find that $H(N)$ annihilates $|\psi_{Diff}\rangle$
which completes the proof.


\begin{references}
\bibitem{Simon} We would like to thank W. Simon for first bringing this
graph to our attention.
\end{references}
\end{document}